\documentclass[11pt]{article}

\usepackage[preprint]{acl}

\usepackage{times}
\usepackage{latexsym}

\usepackage[T1]{fontenc}

\usepackage[utf8]{inputenc}

\usepackage{microtype}

\usepackage{inconsolata}

\usepackage{graphicx}
\usepackage{booktabs}
\usepackage{amsmath}

\title{MM-IssueLoc: A Controlled Benchmark for Evaluating Visual Evidence in Multimodal Repository-Level Issue Localization}

\author{
\textbf{Shaoxiong Zhan}\textsuperscript{1,2}
\thanks{Equal contribution.},
\textbf{Shi Hu}\textsuperscript{2}\footnotemark[1],
\textbf{Boyu Feng}\textsuperscript{2},
\textbf{Hai Lin}\textsuperscript{1},
\textbf{Andrew Gong}\textsuperscript{2},\\
\textbf{Zhengda Zhou}\textsuperscript{2},
\textbf{Jiaying Zhou}\textsuperscript{2},
\textbf{Yunyun Hou}\textsuperscript{2},
\textbf{Hao Su}\textsuperscript{2},
\textbf{Hai-Tao Zheng}\textsuperscript{1}
\thanks{Corresponding author.} \\
\textsuperscript{1}Tsinghua University \quad
\textsuperscript{2}JD.com \\
\texttt{zhansx24@mails.tsinghua.edu.cn},
\texttt{hushi.5@jd.com},
\texttt{zheng.haitao@sz.tsinghua.edu.cn}
}

\usepackage{cleveref}

\begin{document}
\maketitle
\begin{abstract}
  Real repository issues routinely include visual evidence such as screenshots, error dialogs, rendered UI states, and logs, yet repository-level issue localization is evaluated mostly as a text-only task. Existing multimodal SE benchmarks evaluate end-to-end repair, entangling localization with patch synthesis and obscuring whether visual input helped, hurt, or was ignored. We introduce \textbf{MM-IssueLoc}, a controlled benchmark and evaluation protocol for repository-level localization with visual evidence. MM-IssueLoc contains 652 issue-PR instances across 23 languages, with annotations for 7 image categories and 4 relevance levels. It provides file-level and function-level gold labels, paired text-only and with-image evaluation, and VCE-based diagnostics that convert images into structured textual evidence. We evaluate LLM-based and retrieval-based systems, including MM-IssueLoc-VL-Emb as a controlled multimodal retriever. Results show that existing systems remain far from reliable multimodal repository localization: the strongest agent reaches 38.96 file Acc@5 and 22.45 function Acc@10, while the strongest retriever reaches 33.86 function Acc@10. Cross-benchmark comparisons show that high localization scores on text-dominant SWE benchmarks do not transfer cleanly to multimodal issue localization. MM-IssueLoc turns visual evidence into an explicit evaluation variable, enabling future work to test whether systems improve by using visual evidence for localization, rather than by relying on text-only cues or downstream patch-generation effects. \url{https://github.com/Jasaxion/MM-IssueLoc-Bench}
\end{abstract}

\begin{figure*}[t]
  \centering
  \includegraphics[width=0.98\linewidth]{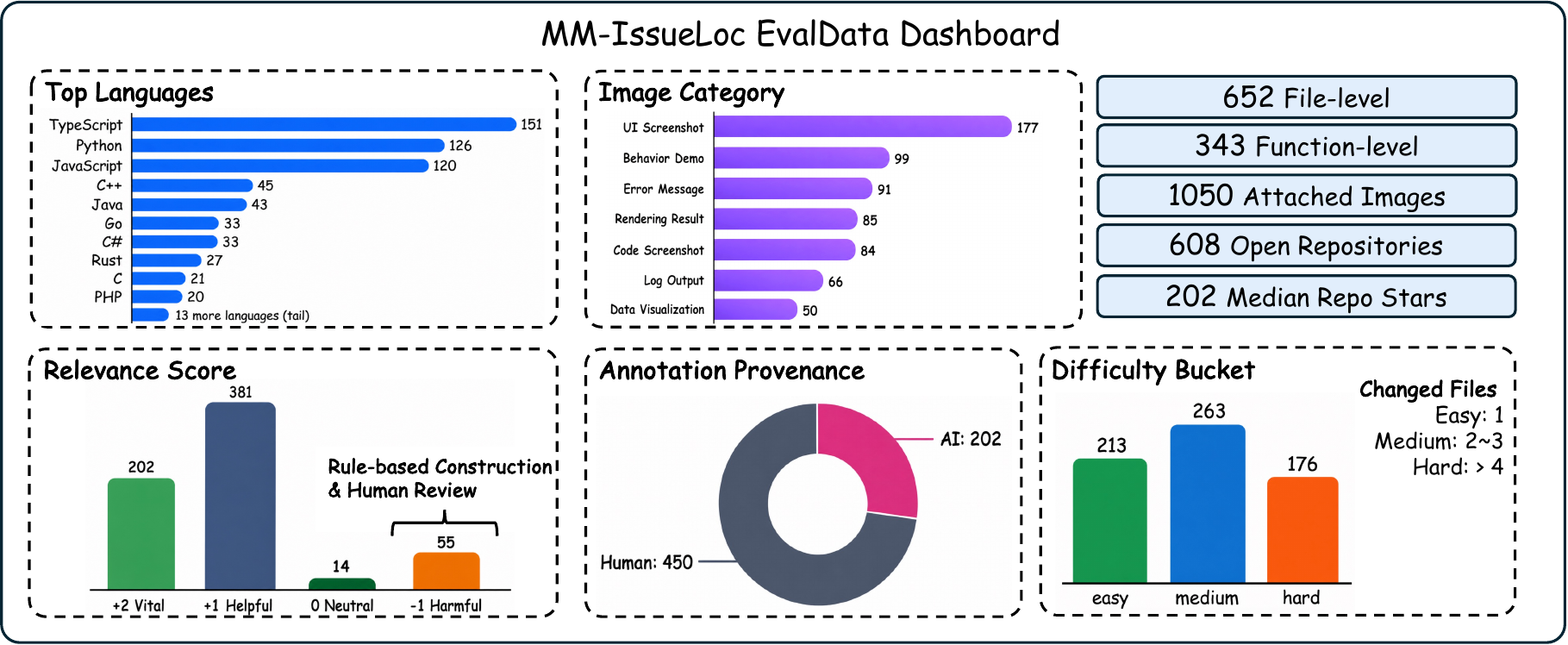}
   \caption{Dashboard of MM-IssueLoc. MM-IssueLoc contains 652 file-level instances and 343 function-level instances, drawn from 608 repositories across 23   
  programming languages, with 1{,}050 attached issue images. Each panel reports the distribution of one annotation dimension: programming     
  language, image category, relevance score, annotation provenance, and difficulty bucket.} 
  \label{fig:bench-stats}
\end{figure*}

\section{Introduction}
\label{sec:intro}

Repository-level issue localization takes an issue description and a repository snapshot as input, and identifies the code locations relevant to resolving the issue~\citep{chen2025locagent}. For automated software engineering (SE) systems, localization is a prerequisite for issue resolving: errors at this step propagate into every downstream stage of patch generation. Real-world repository issues are natively multimodal; developers routinely attach screenshots, UI states, error dialogs, rendered outputs, and log excerpts alongside textual descriptions. Recent work has begun to extend SE benchmarks beyond text, e.g., SWE-bench Multimodal~\citep{yang2024swebench}, which studies whether SE agents generalize to visually grounded tasks.

However, no existing benchmark isolates the role of visual evidence at the localization step. Text-only benchmarks such as SWE-bench and its multilingual extensions strip images entirely~\citep{jimenez2023swebench,zan2025multiswebench}. SWE-bench Multimodal~\citep{yang2024swebench} retains images but evaluates end-to-end patch generation, so localization is folded into repair, and whether an image helped identify the right file or function cannot be read off the final pass/fail signal. Neither setting provides fine-grained localization gold, per-image annotations, or paired with-image versus without-image runs. These are the ingredients needed to treat visual evidence as a first-class input to repository-level localization.

We introduce \textbf{MM-IssueLoc}, a controlled benchmark and evaluation protocol for multimodal repository level issue localization. Each instance is grounded in a real GitHub issue, its linked pull request, and the repository snapshot before the fix. MM-IssueLoc contains 652 issue PR instances across 23 programming languages, with 1{,}050 issue images, file level gold labels for all instances, and function level gold labels for 343 instances. Each image is annotated with one of seven evidence categories, such as UI screenshot, error message, rendering result, code screenshot, log output, and data visualization, as well as a relevance score indicating whether the image is harmful, neutral, helpful, or vital for localization. The benchmark also includes 55 human reviewed harmful image instances, which serve as a robustness stress test for visually plausible but localization incorrect evidence.

MM-IssueLoc is designed as an evaluation instrument rather than as a leaderboard for a single model family. It supports four paired input modes: text-only, with-image, Visual Content Evidence (VCE), and VCE+image. These modes allow us to separate the effect of visual information from the effect of raw image conditioning. We evaluate both llm-based and retrieval-based localization systems, including AgentLess, LocAgent, OpenHands, Mini-SWE-Agent, BM25, SweRank, Qwen3-VL-Embedding, and our controlled multimodal retriever,
MM-IssueLoc-VL-Embedding~\citep{xia2024agentless,chen2025locagent,wang2025openhands,reddy2025swerank,li2026qwen}. We further compare representative systems across SWE-bench Lite, SWE-bench Verified, SWE-bench Multimodal, and MM-IssueLoc (Appendix~\ref{sec:app_cross_bench}). These cross-benchmark results show that strong localization performance on text-dominant SWE benchmarks does not transfer cleanly to multimodal issue localization, which highlights the need for a benchmark that controls visual evidence explicitly.

Our findings show that MM-IssueLoc exposes a capability gap that is difficult to observe in text dominant benchmarks. On MM-IssueLoc, the best file level result reaches 38.96 Acc@5, and the best function level agent result reaches 22.45 Acc@10. Controlled ablations further show that visual evidence can provide useful localization signal: removing images reduces MM-IssueLoc-VL-Embedding by 4.44 to 4.91 File@5 points, while VCE improves several agent framework pairs by converting visual content into structured text. These results motivate MM-IssueLoc as a controlled benchmark for evaluating when and how visual evidence contributes to repository level localization.

Our contributions are three-fold: 1) we introduce \textbf{MM-IssueLoc}, a controlled multimodal issue localization benchmark with 652 real issue PR instances, 1{,}050 images, seven evidence categories, four relevance levels and file/function level gold labels; 2) we provide a controlled evaluation protocol with text only, with image, VCE, and VCE+image modes for isolating visual evidence under strict Acc@K metrics; and 3) we evaluate representative agents and retrievers, introduce MM-IssueLoc-VL-Embedding as a controlled ablation model, and show that visual evidence offers useful but model dependent localization signal.

\section{Related Work}
\label{sec:related_work}


\paragraph{Repository-level SE benchmarks.}
SWE-bench establishe repository issue resolving as a scalable benchmark for automated SE systems~\citep{jimenez2023swebench}. Its later variants and related multilingual efforts broaden the issue-resolving ecosystem, including multilingual coverage in Multi-SWE-bench~\citep{zan2025multiswebench} and larger data-construction efforts such as SWE-smith~\citep{yang2025swesmith}. SWE-bench Multimodal is the closest prior artifact to MM-IssueLoc: it asks whether SE agents generalize to visual, user-facing JavaScript tasks~\citep{yang2024swebench}. However, these benchmarks evaluate issue resolving as a whole. MM-IssueLoc instead isolates localization, adds per-image relevance labels and harmful-image controls, and reports both file-level and function-level gold across 23 languages.

\paragraph{Repository Search and Localization agents.}
Recent SE agents provide different ways for models to inspect and navigate repositories. SWE-agent studies agent-computer interfaces for software engineering~\citep{yang2024sweagent}, Agentless argues that much of end-to-end repair can be decomposed into simpler localization and patching components~\citep{xia2024agentless}, and OpenHands exposes a production-oriented SDK for software agents~\citep{wang2025openhands}. Beyond these general software-agent frameworks, recent work has increasingly focused on specialized repository search. LocAgent and CoSIL formulate localization as an agentic, iterative search process over repository structure, where the model repeatedly inspects files, functions, and contextual evidence before selecting suspicious locations~\citep{chen2025locagent,jiang2025issue}. RepoSearcher and CodeScout instead emphasize learned repository search policies, including reinforcement-learning-based training, to improve how models navigate large codebases and identify relevant code regions~\citep{ma2025tool, sutawika2026codescout}. MM-IssueLoc uses these systems as baselines but evaluates only localization outputs. This design separates localization ability from patch-generation success and from framework-specific submission failures.

\paragraph{Code retrieval and embedding models.}
Retrieval methods offer lower-latency alternatives to full agents. SWERank and SWERank+ formulate issue localization as code ranking, including function-level and multilingual settings~\citep{reddy2025swerank,reddy2025swerankb}. CodeXEmbed and large-scale code representation learning study general-purpose code embeddings~\citep{liu2024codexembed,zhang2024code}. Multimodal embedding models such as Qwen3-VL-Embedding and jina-embeddings-v4 extend dense retrieval to visual and textual inputs~\citep{li2026qwen,gnther2025jinaembeddingsv}. MM-IssueLoc-VL-Embedding builds on this line but serves a different purpose: it is a controlled retriever for testing image-presence, stage-curriculum, and hard-negative design under a fixed benchmark protocol. Its contrastive objective follows the general InfoNCE/CLIP-style retrieval paradigm~\citep{oord2018representation,radford2021learning}.

\paragraph{Multimodal code understanding.}
Recent work has introduced visual evidence into software-engineering evaluation. MMCode studies visually rich programming problems~\citep{li2024mmcode}, while FailureMem and long-horizon multimodal search examine visual evidence in broader repair or search pipelines~\citep{ma2026failuremem,du2026towards}. These studies motivate multimodal software reasoning, but they do not provide a controlled benchmark for repository-level issue localization. MM-IssueLoc complements this line by isolating localization from repair and by evaluating how visual evidence affects file- and function-level localization.

\begin{table}[t]
\centering
\small
\setlength{\tabcolsep}{5pt}
\resizebox{1.0\linewidth}{!}{
\begin{tabular}{lcccc}
\toprule
Benchmark & Multimodal & Repo-level & Image relevance & Function gold \\
\midrule
SWE-bench / Verified & No & Yes & -- & Partial \\
SWE-bench Multimodal & Yes & Yes & No & No \\
Multi-SWE-bench & No & Yes & -- & Patch-derived \\
LocAgent / Loc-Bench & No & Yes & -- & Yes \\
MM-IssueLoc & Yes & Yes & Yes & Yes \\
\bottomrule
\end{tabular}}
\caption{Positioning of MM-IssueLoc. The benchmark is designed to isolate visual evidence in repository-level localization, not to replace end-to-end issue-resolving benchmarks.}
\label{tab:positioning}
\end{table}

\section{Benchmark Construction}
\label{sec:bench}

MM-IssueLoc evaluates whether visual evidence in real issue reports helps localize code responsible for repository-level failures. To make this measurable, each instance is grounded in a GitHub issue, linked pull request, and fixed repository snapshot; supports paired evaluation with and without images; includes per-image evidence type and localization relevance annotations; and provides both file-level and function-level gold labels.

\subsection{Task Definition}
\label{sec:task-definition}

Given an issue report and a repository snapshot, the task is to rank code
locations by their likelihood of being responsible for the reported failure.
Each MM-IssueLoc instance is defined as
\begin{equation}
\label{eq:instance-definition}
x_i = (t_i, b_i, I_i, R_i@c_i).
\end{equation}
where \(t_i\) is the issue title, \(b_i\) is the issue body,
\(I_i=\{i_{i,1},\ldots,i_{i,n}\}\) is a set of attached images, and
\(R_i@c_i\) denotes the repository snapshot at the base commit of the linked
pull request. A system returns either an ordered list of candidate files
\begin{equation}
\label{eq:file-ranking}
\hat{F}_i = [f_{i,1}, \ldots, f_{i,K}],
\end{equation}
for file-level localization, or an ordered list of qualified functions
\begin{equation}
\label{eq:function-ranking}
\hat{G}_i = [g_{i,1}, \ldots, g_{i,K}],
\end{equation}
for function-level localization.

The gold file set \(F_i^\star\) consists of files edited by the human-vetted fixing pull request. For instances with function-level annotations, the gold function set \(G_i^\star\) consists of edited functions that can be matched to functions in the pre-fix repository snapshot \(R_i@c_i\). Functions newly introduced by the pull request are excluded, since they are absent from the input repository and therefore cannot be localized. Function-level gold labels are provided only when tree-sitter extraction succeeds and at least one edited pre-existing function can be identified.

\subsection{Data Collection and Annotation}
\label{sec:data-collection}

MM-IssueLoc is built from public GitHub repositories with at least 50 stars, closed and merged pull requests, non-bot authorship, deduplication, GitHub API enrichment, issue-PR linking, pre/post file retrieval, and complete repository snapshots at the base commit. The released benchmark contains 652 instances, 1,050 localized issue images, and 650 repository snapshots. Its file-level view covers all 652 instances; its function-level view covers 343 instances.

Human annotation covers 450 instances by six annotators over approximately 20 hours. Each instance receives one image category among \emph{ui\_screenshot}, \emph{behavior\_demo}, \emph{error\_message}, \emph{rendering\_result}, \emph{code\_screenshot}, \emph{data\_visualization}, and \emph{log\_output}; a relevance score in $\{-1,0,+1,+2\}$; and a difficulty bucket derived from the number of changed files. The AI-assisted extension adds 202 instances through a four-stage VLM gate: relevance scoring, category classification with confidence threshold, patch-alignment verification, and a second-model consistency check. All AI rows retain \emph{annotation\_by=ai}, which enables separate sanity checks.

This split is audited in the evaluation rather than treated as bookkeeping. On file Acc@5, OpenHands Claude scores 36.00 on the human subset and 37.13 on the AI subset, while AgentLess Claude scores 27.39 and 28.00. The ranking is stable, suggesting that the AI-assisted extension broadens coverage without dominating the benchmark behavior. further benchmark construction details are provided in Appendix~\ref{app:benchmark-details}.

\subsection{Function-Level Gold and Harmful-Image Controls}
\label{sec:function-gold-and-harmful-image-controls}

Function-level labels are extracted by intersecting unified-diff line ranges with tree-sitter AST function spans. Functions newly introduced by the PR are excluded from both gold and candidate pools because the task is localization in the base repository, not prediction of new code. Languages with weak grammar coverage can still be evaluated at file level.

Naturally harmful images are rare. In the human pool, images that actively mislead localization occur at less than 1\%. MM-IssueLoc therefore includes controlled harmful-image synthesis for robustness testing. The synthesis uses three strategies: same-repository image-text mismatch, TF-IDF-similar issues with disjoint edit files, and promotion of misleading comment-thread screenshots into the issue-body image slot. Each synthetic instance is scored along seven dimensions, including text sufficiency, surface relevance, image direction, direction deviation, misleading score, developer impact, and suggested relevance label, followed by human review.

\begin{figure*}[t]
\centering
\includegraphics[width=0.98\linewidth]{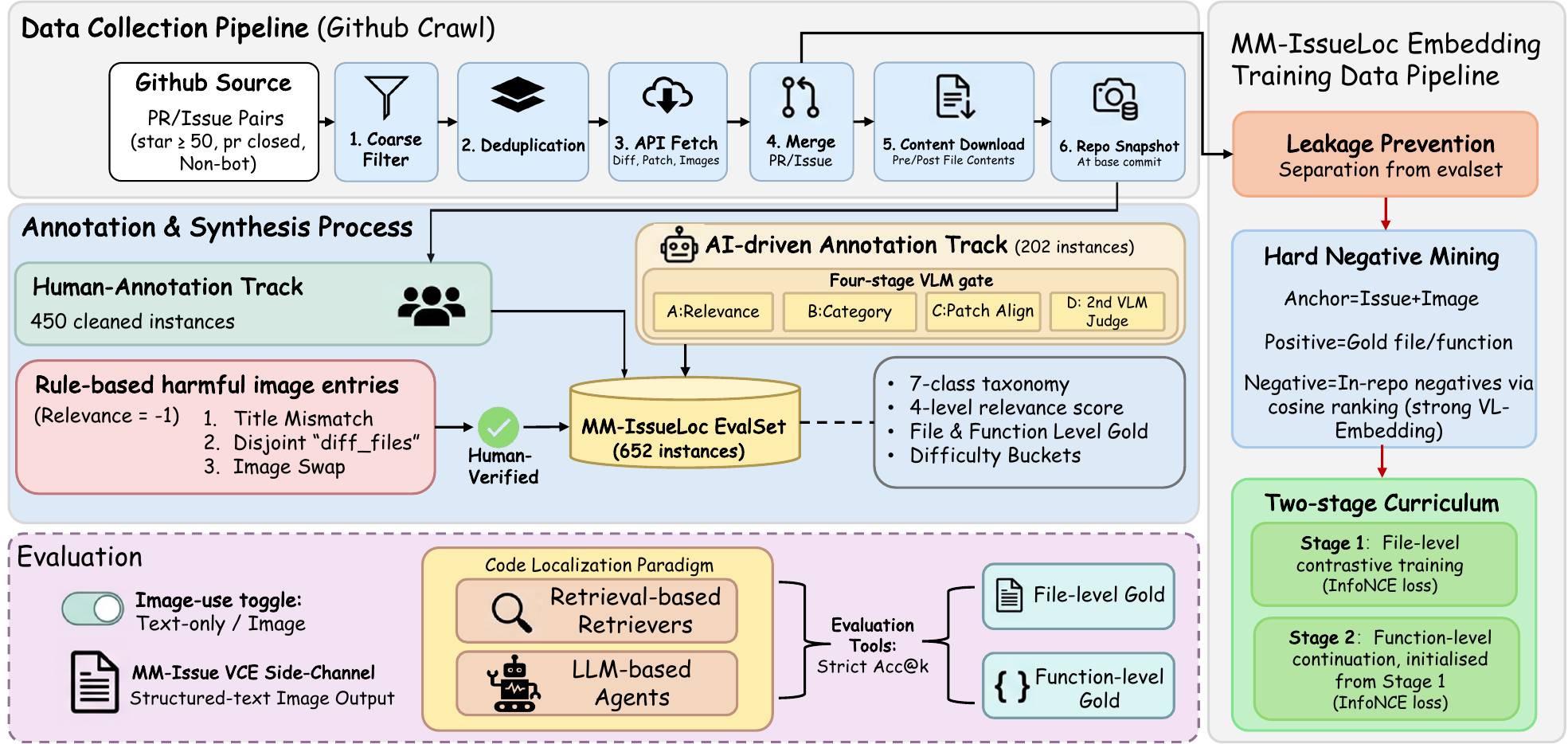}
\caption{End-to-end construction and evaluation flow for MM-IssueLoc. The benchmark separates human and AI annotation, explicit harmful-image controls, file/function-level gold extraction, multimodal training data construction, and controlled image-conditioning evaluation modes.}
\label{fig:pipeline}
\end{figure*}

\subsection{Training Data and Controlled Retriever}
\label{sec:training-data-retriever}

The companion MM-IssueLoc training data is disjoint from the benchmark at the
\texttt{instance\_id} level and follows the right branch of the pipeline in
\Cref{fig:pipeline}. It contains 13,196 file-level rows over 7,664 instances
and 13,769 function-level rows over 3,999 instances. Each anchor is multimodal
and uses at most two images. Multi-positive expansion emits one row per gold
file or function, aligning training with the multi-gold evaluation protocol.
For each anchor, hard negatives~\citep{robinson2021contrastive} are mined from same-repository non-gold
candidates ranked by Qwen3-VL-Embedding-8B, so the model contrasts the true
target with plausible repository-local distractors. MM-IssueLoc-VL-Embedding is
initialized from Qwen3-VL-Embedding-2B or 8B. The query contains issue text and
images; candidates are file or function strings prefixed with \texttt{File:}
or \texttt{Function:}. Let \(q_i\), \(p_i\), and \(n_{ij}\) denote the
normalized query, positive, and hard-negative embeddings. We optimize
\[
\mathcal{L}_i = -\log
\frac{\exp(q_i^\top p_i / \tau)}
{\exp(q_i^\top p_i / \tau) + \sum_j \exp(q_i^\top n_{ij}/\tau)},
\]
with \(\tau=0.05\). Training follows a file-to-function curriculum: Stage 1
fine-tunes on file-level rows, and Stage 2 continues on function-level rows.

Overall, MM-IssueLoc contains 652 issue instances, 1,050 images, and covers 24
programming languages. All instances support file-level evaluation, and 343
also support function-level evaluation. Each instance records annotation
provenance, image evidence category, image relevance score, and difficulty,
enabling controlled analyses by language, visual evidence type, relevance,
annotation source, and search-space size. \Cref{fig:bench-stats} summarizes
these benchmark statistics, and the released JSONL files contain the same
fields used in evaluation, making the statistics directly reproducible.

\section{Experimental Setup}
\label{sec:setup}

\subsection{Baselines}
\label{sec:baselines}

We evaluate two representative paradigms for issue localization: retrieval-based localization and llm-based localization. These baselines cover both direct relevance ranking and interactive repository exploration, allowing us to evaluate whether visual evidence improves localization under different levels of search and reasoning complexity.

\textbf{Retrieval-based Methods}. This paradigm directly ranks repository files or functions based on relevance score, including BM25~\citep{robertson2009probabilistic}, bge-m3~\citep{chen2024bge},
SWERank-small, SWERank-large~\citep{reddy2025swerank}, Qwen3-VL-Embedding-2B/8B~\citep{li2026qwen}, and
MM-IssueLoc-VL-Embedding-2B/8B.

\textbf{Llm-based methods} explore the
repository with tool use before returning localized code locations, including
AgentLess~\citep{xia2024agentless}, LocAgent~\citep{chen2025locagent}, OpenHands~\citep{wang2025openhands}, and Mini-SWE-Agent~\citep{yang2024sweagent} with frontier and open-weight
multimodal models. All methods are scored by the same evaluation harness.
Agents are constrained to output strict JSON for file/function localization and
are not allowed to generate patches or modify repositories.

For multimodal benchmarks, we evaluate four input modes. The \emph{original} mode passes issue text and raw images. The \emph{text-only} mode removes images. The \emph{+VCE} mode replaces images with a structured visual text block. The \emph{+VCE+image} mode passes both the structured text and raw images.

\subsection{Visual Context Extraction (VCE)}
\label{sec:vce}

VCE is a non-trainable image-to-text adapter. For each image, a VLM extracts structured fields such as OCR text, error signals, UI elements, user actions, code hints, saliency, confidence, and notes. The fields are rendered as a compact textual block and appended to the issue body. The text-only to +VCE comparison measures the value of structured visual content, while original to +VCE+image measures whether raw pixels add further signal. See Appendix~\ref{app:vce} for the VCE workflow and case studies.

\subsection{Evaluation Granularities and Metrics}
\label{sec:metrics}

The harness evaluates both file-level (\(n=652\)) and function-level
(\(n=343\)) localization. Following prior localization work~\citep{chen2025locagent}, we use
strict \textbf{Acc@K} as the primary metric:
\begin{equation}
\label{eq:acc-at-k}
\mathrm{Acc@K}
=
\frac{1}{N}\sum_{i=1}^{N}
\mathbf{1}\left[
G_i^\star \subseteq \operatorname{TopK}(r_i)
\right].
\end{equation}
A prediction is successful only if all gold files or qualified functions appear
in the top-\(K\) ranked list. This all-gold criterion is important for
multi-location issues, where finding only one correct location may still fail
to recover the full edit scope. Function-level results are reported at
\(K\in\{5,10\}\), since strict Acc@1 and Acc@3 are necessarily zero whenever an
instance contains more than one gold function. We also report MRR, Recall@K,
and Hit@K as supporting metrics.

\section{Results and Analysis}
\label{sec:results}

\subsection{RQ1: How Well Do Current Systems Localize Multimodal Repository Issues?}

RQ1 establishes the capability boundary of current systems. As shown in
Table~\ref{tab:main_results}, multimodal repository-level issue localization
remains far from solved. The strongest file-level result is obtained by
OpenHands with GPT-5.2, which reaches 38.96 File@5, while OpenHands with
Claude-Sonnet-4.6 achieves the best File@1 and File@3. However, even the best
system still misses all correct files in the top five for more than 60\% of
instances. This suggests that current systems can provide useful candidates,
but are not yet reliable enough for downstream repair-oriented workflows.

The results further reveal a clear paradigm difference. LLM-based agents are
stronger at file-level localization, likely because they can inspect the
repository and follow issue-level clues during exploration. Retrieval-based
methods are more competitive at function-level localization: MM-IssueLoc-VL-Emb-8B
achieves the best Func@10 score of 33.86. This advantage is partly structural.
Retrievers rank pre-enumerated functions that are independently encoded, so
their outputs are naturally aligned with the evaluation space. In contrast,
LLM-based systems must generate function identifiers as text, where incomplete
names, missing qualifiers, or invalid output formats can reduce exact-match
scores.

\begin{table*}[t]
\centering
\scriptsize
\resizebox{0.9\linewidth}{!}{
\begin{tabular}{llrrrrr}
\toprule
\textbf{Paradigm} & \textbf{Model} &
\textbf{File@1} & \textbf{File@3} & \textbf{File@5} &
\textbf{Func@5} & \textbf{Func@10} \\
\midrule
\multicolumn{7}{l}{\textbf{LLM-based localization}} \\
\cmidrule(lr){1-7}
OpenHands & GPT-5.2
& 23.93 & 36.35 & \textbf{38.96} & 21.87 & 22.45 \\
OpenHands & Claude-Sonnet-4.6
& \textbf{25.61} & \textbf{36.66} & 37.42 & 20.41 & 20.41 \\
OpenHands & Kimi-K2.5
& 24.23 & 35.28 & 35.74 & 20.12 & 20.12 \\
AgentLess & Kimi-K2.5
& 18.15 & 28.83 & 32.74 & 12.14 & 12.14 \\
AgentLess & Gemini-3-Pro
& 17.87 & 27.70 & 30.11 & 12.92 & 12.92 \\
LocAgent & Qwen3.5-122B
& 21.36 & 25.39 & 25.70 & 12.98 & 12.98 \\
Mini-SWE-Agent & Claude-Sonnet-4.6
& 15.18 & 18.71 & 18.71 & 10.79 & 11.08 \\
\midrule
\multicolumn{7}{l}{\textbf{Retrieval-based localization}} \\
\cmidrule(lr){1-7}
BM25 & Text
& 6.75 & 16.56 & 21.01 & 12.54 & 19.83 \\
bge-m3-0.5B & Text
& 7.36 & 14.88 & 20.40 & 18.66 & 26.24 \\
SWERank-small-0.1B & Text
& 3.83 & 7.98 & 10.74 & 19.83 & 26.24 \\
SWERank-large-7B & Text
& 10.12 & 21.32 & 26.69 & 20.95 & 27.82 \\
Qwen3-VL-Emb-2B & Multimodal
& 11.66 & 23.16 & 29.29 & 24.20 & 30.90 \\
Qwen3-VL-Emb-8B & Multimodal
& 12.11 & 26.69 & 32.06 & \textbf{25.36} & 32.65 \\
MM-IssueLoc-VL-Emb-2B & Multimodal
& 13.19 & 25.61 & 32.06 & \textbf{25.36} & 32.36 \\
MM-IssueLoc-VL-Emb-8B & Multimodal
& \underline{13.96} & \underline{27.61} & \underline{32.82} & 22.74 & \textbf{33.86} \\
\bottomrule
\end{tabular}}
\caption{Overall comparison on MM-IssueLoc. File metrics are Acc@1/3/5, and function
metrics are Acc@5/10. Values are percentages. Bold numbers indicate the best
result in each column, and underlined numbers indicate the best retrieval-based
file-level results.}
\label{tab:main_results}
\end{table*}

\begin{figure*}[t]
\centering
\includegraphics[width=0.98\linewidth]{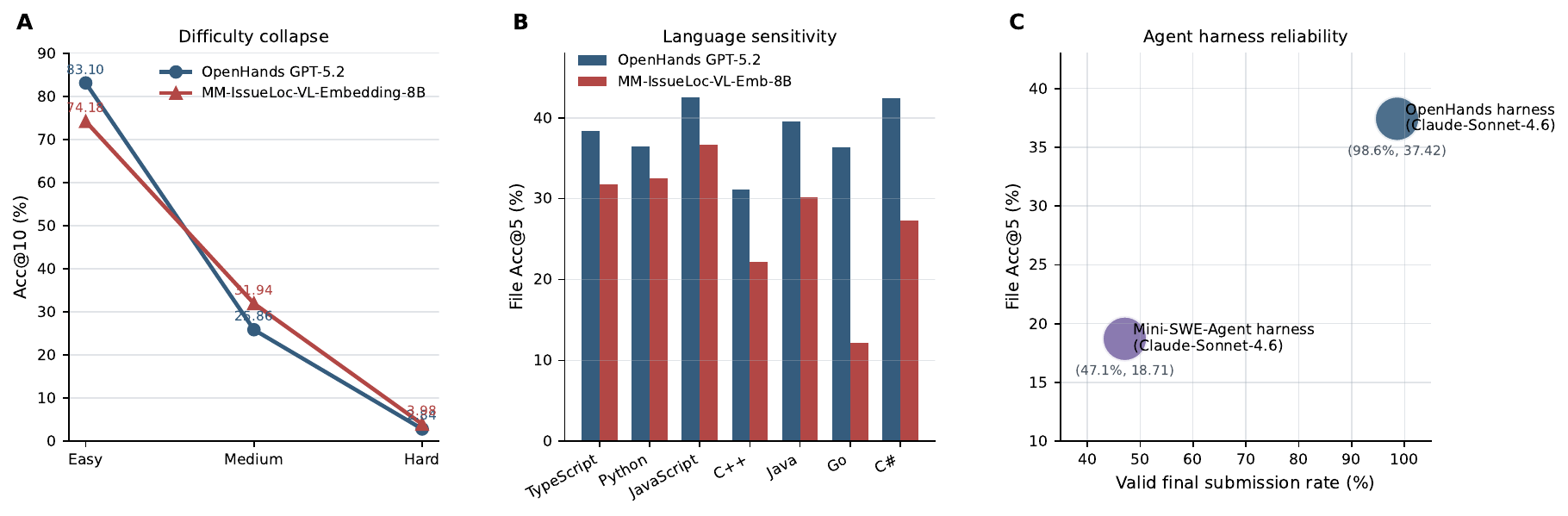}
\caption{Current capability frontier on MM-IssueLoc. Agents are stronger at broad file discovery, retrievers are competitive at function ranking, hard instances collapse across paradigms, and language slices expose additional sensitivity that aggregate scores obscure.}
\label{fig:rq1_capability}
\end{figure*}

Figure~\ref{fig:rq1_capability} further explains the aggregate results. First,
performance collapses with difficulty. OpenHands GPT-5.2 drops from 83.10
Acc@10 on easy instances to 2.84 on hard instances, while
MM-IssueLoc-VL-Emb-8B drops from 74.18 to 3.98, showing that multi-edit issues
remain the main bottleneck. Second, language slices reveal non-uniform
localization behavior. Both systems perform better on JavaScript and related
frontend languages, where screenshots often provide direct UI or rendering
evidence, but performance is weaker on languages such as C++ and Go, where the
visual signal is less directly tied to localized code. Third, agent results are
also shaped by harness reliability. With the same Claude-Sonnet-4.6 backend,
OpenHands achieves a 98.6\% valid submission rate and 37.42 File@5, whereas
Mini-SWE-Agent achieves 47.1\% and 18.71. Thus, RQ1 shows that current systems
are limited not only by model capability, but also by difficulty, language
properties, and final-output reliability.

\begin{table*}[t]
\centering
\small
\setlength{\tabcolsep}{4pt}
\resizebox{0.9\linewidth}{!}{
\begin{tabular}{llrrrr}
\toprule
Method & Model & No-image File@5 & With-image File@5 & $\Delta$ File@5 \\
\midrule
AgentLess & GPT-5.2 & 26.93 & 29.59 & -2.66 \\
OpenHands & GPT-5.2 & 38.65 & 38.96 & -0.31 \\
OpenHands & Claude-Sonnet-4.6 & 38.34 & 37.42 & +0.92 \\
Retriever & MM-IssueLoc-VL-Embed-2B & 27.15 & 32.06 & -4.91\\
Retriever & MM-IssueLoc-VL-Embed-8B & 28.38 & 32.82 & -4.44\\
\bottomrule
\end{tabular}}
\caption{With-image versus no-image ablation on MM-IssueLoc. $\Delta$ is no-image minus with-image file Acc@5. Negative values mean that removing images hurt the cell.}
\label{tab:image_ablation}
\end{table*}

\subsection{RQ2: Is Visual Evidence Useful, and Do Systems Use It Reliably?}

RQ2 separates visual evidence from visual use. MM-IssueLoc is designed for this
distinction: each image is annotated with evidence category and relevance, and
the harmful subset introduces visually plausible but localization-incorrect
cues. Therefore, the no-image ablation evaluates whether systems exploit
image-conditioned localization evidence, rather than whether images are merely
attached to issue reports.

Cross-benchmark results further motivate this question. As reported in
Appendix~\ref{sec:app_cross_bench}, current systems already achieve strong
localization performance on text-dominant SWE benchmarks. For example,
OpenHands reaches 94.53 File@5 on SWE-bench-Lite and 90.20 File@5 on
SWE-bench-Verified. However, performance drops substantially on SWE-bench-MM,
where the best File@5 is 43.14. This gap suggests that repository-level
localization is relatively mature in text-only settings, but remains much less
developed when issue evidence involves images. MM-IssueLoc fills this diagnostic
gap by providing controlled image relevance, category, and harmful-image
annotations.

The main results also indicate that MM-IssueLoc contains usable visual signal.
In Table~\ref{tab:main_results}, text-only retrievers reach 19.83--27.82
Func@10, whereas MM-IssueLoc-VL-Embedding-8B achieves 33.86 Func@10. This shows
that visual evidence can be converted into fine-grained localization signal when
a retriever is trained for multimodal issue localization. Detailed retriever
ablations are reported in Appendix~\ref{sec:app_retriever_ablations}.

Table~\ref{tab:image_ablation} reports the controlled no-image ablation. We define
\[
\begin{aligned}
\Delta_{\mathrm{img}}(m)
={}& \mathrm{Acc@5}_{\mathrm{file}}
      (m,\text{text-only}) \\
 &- \mathrm{Acc@5}_{\mathrm{file}}
      (m,\text{text+image}),
\end{aligned}
\]
so negative values indicate that removing images hurts. The results show uneven image use across systems. AgentLess GPT-5.2 benefits from images
($\Delta_{\mathrm{img}}=-2.66$), while OpenHands changes by less than one point.
MM-IssueLoc-VL-Embedding is more image-sensitive: removing images reduces
File@5 by 4.91 points for the 2B model and 4.44 points for the 8B model. Thus,
visual evidence is useful, but current agent systems do not exploit it
consistently.

Fine-grained slices explain why aggregate image gains are unstable. Relevance
labels do not induce monotonic accuracy changes, since harmful images are not
always recognized as misleading and some instances remain solvable from text
alone. Category slices further show that images are most helpful when they are
directly tied to observable software behavior, such as UI screenshots, behavior
demos, rendering results, code screenshots, logs, or data visualizations. These
signals often help identify relevant files, but they do not always map cleanly
to edited functions. Additional relevance, category, difficulty, and annotation
slices are reported in Appendix~\ref{sec:app_stratified_slices}.

\subsection{RQ3: Does Localization Depend on Image Content or Raw Image Format?}

RQ3 diagnoses whether the benefit of visual evidence comes from image content or from the raw image format. We use Visual Content Evidence (VCE)(\cref{app:vce_method_overview}) to convert each image into structured textual evidence. The +VCE setting removes raw images and provides only structured visual text, while the +VCE+image setting provides both VCE and the original image. Thus, the gain from the original setting to +VCE indicates whether image content is useful after being textualized, and the difference between +VCE and +VCE+image measures the residual value of raw pixels.

Table~\ref{tab:vce} reports the VCE diagnostic results using the same three model backends under AgentLess and LocAgent. VCE improves file-level localization in several settings. For AgentLess, Claude-Sonnet-4.6 improves from 25.15 original File@5 to 28.53 with +VCE, and GPT-5.2 improves from 28.15 to 29.07. For LocAgent, Qwen3.5-122B-A10B improves substantially from 25.70 to 35.04. These results show that images contain useful localization evidence and that such evidence can be consumed in structured textual form.

However, raw images do not consistently add value once VCE is provided.
We define the raw-pixel residual as
\[
\rho_{\mathrm{pix}} =
\mathrm{Acc@5}_{f}(\text{VCE+img}) -
\mathrm{Acc@5}_{f}(\text{VCE}).
\]
For AgentLess, the residuals are small: $+0.22$ for Claude-Sonnet-4.6, $+0.03$ for GPT-5.2, and $-2.15$ for Qwen3.5-122B-A10B. In contrast, LocAgent exhibits much larger variation: Claude-Sonnet-4.6 drops by 5.62 points, GPT-5.2 gains 4.74 points, and Qwen3.5-122B-A10B drops by 11.69 points. This indicates that VCE exposes useful image content, while the effect of raw pixels is mediated by the framework's visual-input integration.

\begin{table*}[t]
\centering
\small
\resizebox{0.7\linewidth}{!}{
\begin{tabular}{llrrrr}
\toprule
\textbf{Method} & \textbf{Model} &
\textbf{Original} &
\textbf{+VCE} &
\textbf{+VCE+Image} &
\textbf{$\rho_{\mathrm{pix}}$} \\
\midrule
AgentLess & Claude-Sonnet-4.6 & 25.15 & 28.53 & 28.75 & 0.22 \\
AgentLess & GPT-5.2 & 28.15 & 29.07 & 29.10 & 0.03 \\
AgentLess & Qwen3.5-122B-A10B & 28.07 & 26.53 & 24.38 & -2.15 \\
\midrule
LocAgent & Claude-Sonnet-4.6 & 27.73 & 30.93 & 25.31 & -5.62 \\
LocAgent & GPT-5.2 & 19.13 & 16.12 & 20.86 & 4.74 \\
LocAgent & Qwen3.5-122B-A10B & 25.70 & 35.04 & 23.35 & -11.69 \\
\bottomrule
\end{tabular}}
\caption{VCE diagnostic results on MM-IssueLoc, measured by file-level Acc@5. VCE converts image content into structured text, while $\rho_{\mathrm{pix}}$ measures the residual value of raw pixels after VCE has already been supplied.}
\label{tab:vce}
\end{table*}

\section{Discussion, Limitations, and Future Work}
\label{sec:discussion-limitations-future}

\paragraph{Discussion.}
MM-IssueLoc follows prior issue-resolving benchmarks by using merged pull
request diffs as localization gold labels. Although such labels are not minimal
causal explanations, they provide a scalable and auditable proxy for code
locations modified by human developers. We therefore use strict all-gold Acc@K
to measure whether a method recovers the full edit scope, and provide
function-level labels when tree-sitter can reliably map edits to pre-existing
functions. We also release per-instance predictions to support reanalysis under
alternative scoring rules.

\paragraph{Limitations.}
MM-IssueLoc is designed for controlled diagnosis, and its results should be
interpreted within this scope. The harmful-image subset is primarily
rule-constructed and human-reviewed, so it serves as a robustness stress test
rather than an estimate of real-world frequency. VCE is a diagnostic protocol,
not an upper bound on visual reasoning, since it uses fixed extraction fields
and a fixed downstream localization setting. Function-level labels may be noisy
for language features that are difficult for tree-sitter to resolve, such as
macros, templates, or generated code. Our retriever also uses at most two images
per issue, leaving many-image issues as a future stress case. Finally, MM-IssueLoc evaluates localization only, rather than end-to-end
software repair or downstream patch generation.

\paragraph{Future work.}
Future work can extend MM-IssueLoc along three directions: collecting more
naturally occurring misleading images and many-image issues, improving
function-level annotation with language-specific parsers or hybrid static
analysis, and studying whether better multimodal localization improves broader
software-engineering workflows such as repository understanding, debugging,
triage, and downstream repair.

\section{Conclusion}
\label{sec:conclusion}

MM-IssueLoc shows that multimodal repository-level issue localization is a distinct and under-evaluated capability. Cross-benchmark results indicate that strong localization performance on text-dominant SWE benchmarks is not sufficient evidence of multimodal issue localization ability. By pairing real issue-PR instances with file-level and function-level gold labels, image-category and relevance annotations, and controlled text-only, with-image, VCE, and VCE+image modes, MM-IssueLoc makes visual evidence an explicit evaluation variable. Our results show that visual evidence can provide useful localization signal, but current agents and retrievers use it unevenly. Overall, MM-IssueLoc provides a controlled basis for evaluating how visual evidence affects repository-level localization.

\bibliography{custom}

@inproceedings{jimenez2023swebench,
  title={SWE-bench: Can Language Models Resolve Real-World GitHub Issues?},
  author={Jimenez, Carlos E and Yang, John and Wettig, Alexander and Yao, Shunyu and Pei, Kexin and Press, Ofir and Narasimhan, Karthik R},
  booktitle={The twelfth international conference on learning representations},
  year={2023}
}

@inproceedings{yang2024swebench,
  title={SWE-bench Multimodal: Do AI Systems Generalize to Visual Software Domains?},
  author={Yang, John and Jimenez, Carlos E and Zhang, Alex L and Lieret, Kilian and Yang, Joyce and Wu, Xindi and Press, Ori and Muennighoff, Niklas and Synnaeve, Gabriel and Narasimhan, Karthik R and others},
  year     = {2024},
  booktitle={The Thirteenth International Conference on Learning Representations}
}

@inproceedings{zan2025multiswebench,
  title={Multi-SWE-bench: A Multilingual Benchmark for Issue Resolving},
  author={Zan, Daoguang and Huang, Zhirong and Liu, Wei and Chen, Hanwu and Xin, Shulin and Zhang, Linhao and Liu, Qi and Li, Aoyan and Chen, Lu and Zhong, Xiaojian and others},
  year     = {2025},
  booktitle={The Thirty-ninth Annual Conference on Neural Information Processing Systems Datasets and Benchmarks Track}
}

@article{xia2024agentless,
  title={Agentless: Demystifying LLM-based Software Engineering Agents},
  author={Xia, Chunqiu Steven and Deng, Yinlin and Dunn, Soren and Zhang, Lingming},
  journal={arXiv preprint arXiv:2407.01489},
  year={2024}
}

@article{yang2024sweagent,
  title={SWE-agent: Agent-Computer Interfaces Enable Automated Software Engineering},
  author={Yang, John and Jimenez, Carlos E and Wettig, Alexander and Lieret, Kilian and Yao, Shunyu and Narasimhan, Karthik and Press, Ofir},
  journal={Advances in Neural Information Processing Systems},
  volume={37},
  pages={50528--50652},
  year={2024}
}

@article{wang2025openhands,
  title={The OpenHands Software Agent SDK: A Composable and Extensible Foundation for Production Agents},
  author={Wang, Xingyao and Rosenberg, Simon and Michelini, Juan and Smith, Calvin and Tran, Hoang and Nyst, Engel and Malhotra, Rohit and Zhou, Xuhui and Chen, Valerie and Brennan, Robert and others},
  journal={arXiv preprint arXiv:2511.03690},
  year={2025}
}

@inproceedings{chen2025locagent,
  title={LocAgent: Graph-Guided LLM Agents for Code Localization},
  author={Chen, Zhaoling and Tang, Robert and Deng, Gangda and Wu, Fang and Wu, Jialong and Jiang, Zhiwei and Prasanna, Viktor and Cohan, Arman and Wang, Xingyao},
  booktitle={Proceedings of the 63rd Annual Meeting of the Association for Computational Linguistics (Volume 1: Long Papers)},
  pages={8697--8727},
  year={2025}
}

@inproceedings{jiang2025issue,
  title={Issue Localization via LLM-Driven Iterative Code Graph Searching},
  author={Jiang, Zhonghao and Ren, Xiaoxue and Yan, Meng and Jiang, Wei and Li, Yong and Liu, Zhongxin},
  booktitle={2025 40th IEEE/ACM International Conference on Automated Software Engineering (ASE)},
  pages={3034--3045},
  year={2025},
  organization={IEEE}
}

@article{reddy2025swerank,
  title={SweRank: Software Issue Localization with Code Ranking},
  author={Reddy, Revanth Gangi and Suresh, Tarun and Doo, JaeHyeok and Liu, Ye and Nguyen, Xuan Phi and Zhou, Yingbo and Yavuz, Semih and Xiong, Caiming and Ji, Heng and Joty, Shafiq},
  journal={arXiv preprint arXiv:2505.07849},
  year={2025}
}

@article{reddy2025swerankb,
  title={SweRank+: Multilingual, Multi-Turn Code Ranking for Software Issue Localization},
  author={Gangi Reddy, Revanth and Liu, Ye and Zhao, Wenting and Doo, JaeHyeok and Suresh, Tarun and Lee, Daniel and Xiong, Caiming and Zhou, Yingbo and Yavuz, Semih and Joty, Shafiq},
  journal={arXiv e-prints},
  pages={arXiv--2512},
  year={2025}
}

@article{liu2024codexembed,
  title={CodeXEmbed: A Generalist Embedding Model Family for Multiligual and Multi-task Code Retrieval},
  author={Liu, Ye and Meng, Rui and Joty, Shafiq and Savarese, Silvio and Xiong, Caiming and Zhou, Yingbo and Yavuz, Semih},
  journal={arXiv preprint arXiv:2411.12644},
  year={2024}
}

@article{li2026qwen,
  title={Qwen3-VL-Embedding and Qwen3-VL-Reranker: A Unified Framework for State-of-the-Art Multimodal Retrieval and Ranking},
  author={Li, Mingxin and Zhang, Yanzhao and Long, Dingkun and Chen, Keqin and Song, Sibo and Bai, Shuai and Yang, Zhibo and Xie, Pengjun and Yang, An and Liu, Dayiheng and others},
  journal={arXiv preprint arXiv:2601.04720},
  year={2026}
}

@inproceedings{gnther2025jinaembeddingsv,
  title={jina-embeddings-v4: Universal Embeddings for Multimodal Multilingual Retrieval},
  author={G{\"u}nther, Michael and Sturua, Saba and Akram, Mohammad Kalim and Mohr, Isabelle and Ungureanu, Andrei and Wang, Bo and Eslami, Sedigheh and Martens, Scott and Werk, Maximilian and Wang, Nan and others},
  booktitle={Proceedings of the 5th Workshop on Multilingual Representation Learning (MRL 2025)},
  pages={531--550},
  year={2025}
}

@article{zhang2024code,
  title={Code Representation Learning At Scale},
  author={Zhang, Dejiao and Ahmad, Wasi and Tan, Ming and Ding, Hantian and Nallapati, Ramesh and Roth, Dan and Ma, Xiaofei and Xiang, Bing},
  journal={arXiv preprint arXiv:2402.01935},
  year={2024}
}

@inproceedings{li2024mmcode,
  title={MMCode: Benchmarking Multimodal Large Language Models for Code Generation with Visually Rich Programming Problems},
  author={Li, Kaixin and Tian, Yuchen and Hu, Qisheng and Luo, Ziyang and Huang, Zhiyong and Ma, Jing},
  booktitle={Findings of the Association for Computational Linguistics: EMNLP 2024},
  pages={736--783},
  year={2024}
}

@article{ma2026failuremem,
  title={FailureMem: A Failure-Aware Multimodal Framework for Autonomous Software Repair},
  author={Ma, Ruize and Jiang, Yilei and Zhang, Shilin and Ma, Zheng and Feng, Yi and Ng, Vincent and Wang, Zhi and Yue, Xiangyu and Li, Chuanyi and Lu, Lewei},
  journal={arXiv preprint arXiv:2603.17826},
  year={2026}
}

@article{du2026towards,
  title={Towards Long-horizon Agentic Multimodal Search},
  author={Du, Yifan and Liu, Zikang and Peng, Jinbiao and Wu, Jie and Li, Junyi and Li, Jinyang and Zhao, Wayne Xin and Wen, Ji-Rong},
  journal={arXiv preprint arXiv:2604.12890},
  year={2026}
}

@inproceedings{radford2021learning,
  title={Learning Transferable Visual Models From Natural Language Supervision},
  author={Radford, Alec and Kim, Jong Wook and Hallacy, Chris and Ramesh, Aditya and Goh, Gabriel and Agarwal, Sandhini and Sastry, Girish and Askell, Amanda and Mishkin, Pamela and Clark, Jack and others},
  booktitle={International conference on machine learning},
  pages={8748--8763},
  year={2021},
  organization={PmLR}
}

@article{oord2018representation,
  title={Representation Learning with Contrastive Predictive Coding},
  author={Oord, Aaron van den and Li, Yazhe and Vinyals, Oriol},
  journal={arXiv preprint arXiv:1807.03748},
  year={2018}
}

@inproceedings{yang2025swesmith,
  title={SWE-smith: Scaling Data for Software Engineering Agents},
  author={Yang, John and Lieret, Kilian and Jimenez, Carlos E and Wettig, Alexander and Khandpur, Kabir and Zhang, Yanzhe and Hui, Binyuan and Press, Ofir and Schmidt, Ludwig and Yang, Diyi},
  year     = {2025},
  booktitle={The Thirty-ninth Annual Conference on Neural Information Processing Systems Datasets and Benchmarks Track}
}

@inproceedings{robinson2021contrastive,
  title={Contrastive Learning with Hard Negative Samples},
  author={Robinson, Joshua David and Chuang, Ching-Yao and Sra, Suvrit and Jegelka, Stefanie},
  year={2021},
  booktitle={International Conference on Learning Representations}
}

@article{sutawika2026codescout,
  title={CodeScout: An Effective Recipe for Reinforcement Learning of Code Search Agents},
  author={Sutawika, Lintang and Soni, Aditya Bharat and Gandhi, Apurva and Yassine, Taha and Vijayvargiya, Sanidhya and Li, Yuchen and Zhou, Xuhui and Zhang, Yilin and Maben, Leander Melroy and Neubig, Graham and others},
  journal={arXiv preprint arXiv:2603.17829},
  year={2026}
}

@article{ma2025tool,
  title={Tool-Integrated Reinforcement Learning for Repo Deep Search},
  author={Ma, Zexiong and Peng, Chao and Zeng, Qunhong and Gao, Pengfei and Zou, Yanzhen and Xie, Bing},
  journal={arXiv preprint arXiv:2508.03012},
  year={2025}
}

@book{robertson2009probabilistic,
  title={The probabilistic relevance framework: BM25 and beyond},
  author={Robertson, Stephen and Zaragoza, Hugo},
  volume={4},
  year={2009},
  publisher={Now Publishers Inc}
}

@article{chen2024bge,
  title={Bge m3-embedding: Multi-lingual, multi-functionality, multi-granularity text embeddings through self-knowledge distillation},
  author={Chen, Jianlv and Xiao, Shitao and Zhang, Peitian and Luo, Kun and Lian, Defu and Liu, Zheng},
  journal={arXiv preprint arXiv:2402.03216},
  volume={4},
  number={5},
  year={2024}
}

\appendix

\section{Training and Evaluation Setup}
\label{app:setup}

\subsection{Training Hyperparameters}
\label{app:training-hyperparameters}

We report the training details of MM-IssueLoc-VL-Embedding in the multimodal setting, where each input contains the issue text and attached images. Table~\ref{tab:training_config} summarizes the two-stage protocol and the function-only ablation. Stage 1 uses file-level supervision for coarse localization, while Stage 2 continues from the Stage-1 checkpoint with function-level supervision. The function-only ablation directly trains on function-level data to assess the effect of file-level initialization.

\begin{table*}[htbp]
\centering
\small
\caption{Training configurations for the proposed two-stage training and function-only ablation.}
\label{tab:training_config}
\begin{tabular}{lccccc}
\toprule
Setting & Data level & Initialization & Epochs & LR & Seed \\
\midrule
Stage-1 File & File-level & Qwen3-VL-Embedding & 2 & $1\times10^{-5}$ & 42 \\
Stage-2 Function & Function-level & Stage-1 checkpoint & 3 & $5\times10^{-6}$ & 43 \\
Function-only Ablation & Function-level & Qwen3-VL-Embedding & 3 & $1\times10^{-5}$ & 42 \\
\bottomrule
\end{tabular}
\end{table*}

Table~\ref{tab:shared_hyperparams} lists the shared optimization settings. We use full fine-tuning with the vision encoder frozen to reduce training cost while preserving pretrained visual representations.

\begin{table}[htbp]
\centering
\small
\caption{Shared optimization hyperparameters.}
\label{tab:shared_hyperparams}
\resizebox{0.8\linewidth}{!}{
\begin{tabular}{lc}
\toprule
Hyperparameter & Value \\
\midrule
Max sequence length & 8192 \\
Contrastive temperature & 0.05 \\
Warmup ratio & 0.05 \\
Logging steps & 10 \\
Checkpoint interval & 200 steps \\
Vision encoder & Frozen \\
Training strategy & Full fine-tuning \\
Image input in MM setting & Enabled \\
Image input in text-only setting & Disabled \\
\bottomrule
\end{tabular}}
\end{table}

Table~\ref{tab:batch_config} reports the hardware and batch configurations. All experiments use four NVIDIA H200 GPUs. The effective batch size equals the per-GPU batch size multiplied by gradient accumulation steps and the number of GPUs.

\begin{table}[htbp]
\centering
\small
\caption{Hardware and batch configurations for different model sizes.}
\label{tab:batch_config}
\resizebox{0.95\linewidth}{!}{
\begin{tabular}{lcccc}
\toprule
Model & GPUs & Batch/GPU & Accum. Steps & Global Batch \\
\midrule
2B & 4$\times$H200 & 4 & 8 & 128 \\
8B & 4$\times$H200 & 1 & 16 & 64 \\
\bottomrule
\end{tabular}
}
\end{table}

\subsection{Agent Evaluation Parameters}
\label{app:agent-eval-params}                                                                                                   
                                        
We configure each agent framework following the setup reported in its original paper, and restrict every framework to localization-only output on a shared evaluation
harness.                                
                                                                                                                                                                        
OpenHands uses a CodeAct-style agent equipped with a terminal, a file editor, and an atomic \texttt{finish(message=...)} submission action. Each run is capped at 10        
iterations, with a follow-up ``produce final JSON now'' nudge issued when no answer is emitted. Mini-SWE-Agent uses a bash-only interface, a 10-step limit, and a           
heredoc-to-file final submission protocol. Both frameworks share the same repository snapshots, scoring code, retry policy for transient failures, and localization-only    
output schema. The observed gap between OpenHands and Mini-SWE-Agent is therefore interpreted as a framework-level submission and tool-surface effect, rather than a pure   
model-capability difference.

\section{Extended Results}

\subsection{Stratified Evaluation Slices}
\label{sec:app_stratified_slices}

Table~\ref{tab:app_relevance_slice} reports the relevance slice used in RQ2. The non-monotonic pattern is visible across systems: vital images are not always easier than helpful images, and harmful images are not automatically detected as misleading.

\begin{table}[h]
\centering
\small
\setlength{\tabcolsep}{5pt}
\resizebox{1.0\linewidth}{!}{
\begin{tabular}{lrrrr}
\toprule
System & Vital & Helpful & Neutral & Harmful \\
\midrule
OpenHands GPT-5.2 & 36.14 & 40.16 & 21.43 & 34.55 \\
Qwen3-VL-Embed-8B & 35.15 & 29.66 & 21.43 & 32.73 \\
MM-IssueLoc-VL-Emb-8B & 30.20 & 32.28 & 28.57 & 34.55 \\
\bottomrule
\end{tabular}}
\caption{File Acc@5 by image relevance score. Harmful rows are a robustness stress test and should not be interpreted as a natural-frequency estimate.}
\label{tab:app_relevance_slice}
\end{table}

Table~\ref{tab:app_category_slice} reports selected category slices. Text-rich categories such as error-message and code screenshots are easier to convert into localization cues than behavior demos or visualizations, but file-level and function-level trends can diverge.

\begin{table}[t]
\centering
\small
\setlength{\tabcolsep}{5pt}
\renewcommand{\arraystretch}{1.08}
\resizebox{1.0\linewidth}{!}{
\begin{tabular}{lccc}
\toprule
\textbf{Category} &
\textbf{OpenHands GPT-5.2} &
\textbf{Qwen3-VL-Emb-8B} &
\textbf{MM-IssueLoc-VL-Emb-8B} \\
\midrule
UI screenshot       & 39.55 & 31.07 & 30.51 \\
Behavior demo       & 40.40 & 40.40 & 36.36 \\
Rendering result    & 40.00 & 24.71 & 25.88 \\
Code screenshot     & 36.90 & 33.33 & 36.90 \\
Log output          & 37.88 & 27.27 & 30.30 \\
Data visualization  & 32.00 & 20.00 & 28.00 \\
\bottomrule
\end{tabular}}
\caption{Representative category slices on MM-IssueLoc. Values are File Acc@5
percentages. The selected categories illustrate that visual evidence is most
useful when it is directly tied to UI behavior, rendering results, code
screenshots, logs, or visualized outputs.}
\label{tab:app_category_slice}
\end{table}

Tables~\ref{tab:app_language_annotation_slice} and~\ref{tab:app_annotation_slice} give the language and annotation-provenance checks referenced in the main text. The human and AI-assisted subsets preserve the same broad ordering for representative agent baselines, while language slices reveal larger variation for retrievers.

\begin{table}[h]
\centering
\small
\setlength{\tabcolsep}{3pt}
\resizebox{\linewidth}{!}{
\begin{tabular}{lrrrrrrrr}
\toprule
System & TypeScript & Python & JavaScript & C++ & Java & Go & C\# & Rust \\
\midrule
OpenHands GPT-5.2 & 38.41 & 36.51 & 42.50 & 31.11 & 39.53 & 36.36 & 42.42 & 40.74 \\
MM-IssueLoc-VL-Emb-8B & 31.79 & 32.54 & 36.67 & 22.22 & 30.23 & 12.12 & 27.27 & 40.74 \\
\bottomrule
\end{tabular}}
\caption{File Acc@5 by programming language for two representative systems. C\# has no function-level gold in the released function subset.}
\label{tab:app_language_annotation_slice}
\end{table}

\begin{table}[h]
\centering
\small
\resizebox{\linewidth}{!}{
\begin{tabular}{lrr}
\toprule
System & Human File@5 & AI-annotation File@5 \\
\midrule
OpenHands Claude-Sonnet-4.6 & 36.00 & 37.13 \\
AgentLess Claude-Sonnet-4.6 & 27.39 & 28.00 \\
\bottomrule
\end{tabular}}
\caption{Annotation-provenance sanity check. The AI-annotation extension does not dominate the benchmark behavior for representative agent baselines.}
\label{tab:app_annotation_slice}
\end{table}

\subsection{Controlled Retriever Training Ablations}
\label{sec:app_retriever_ablations}

\begin{figure}[h]
\centering
\includegraphics[width=0.92\linewidth]{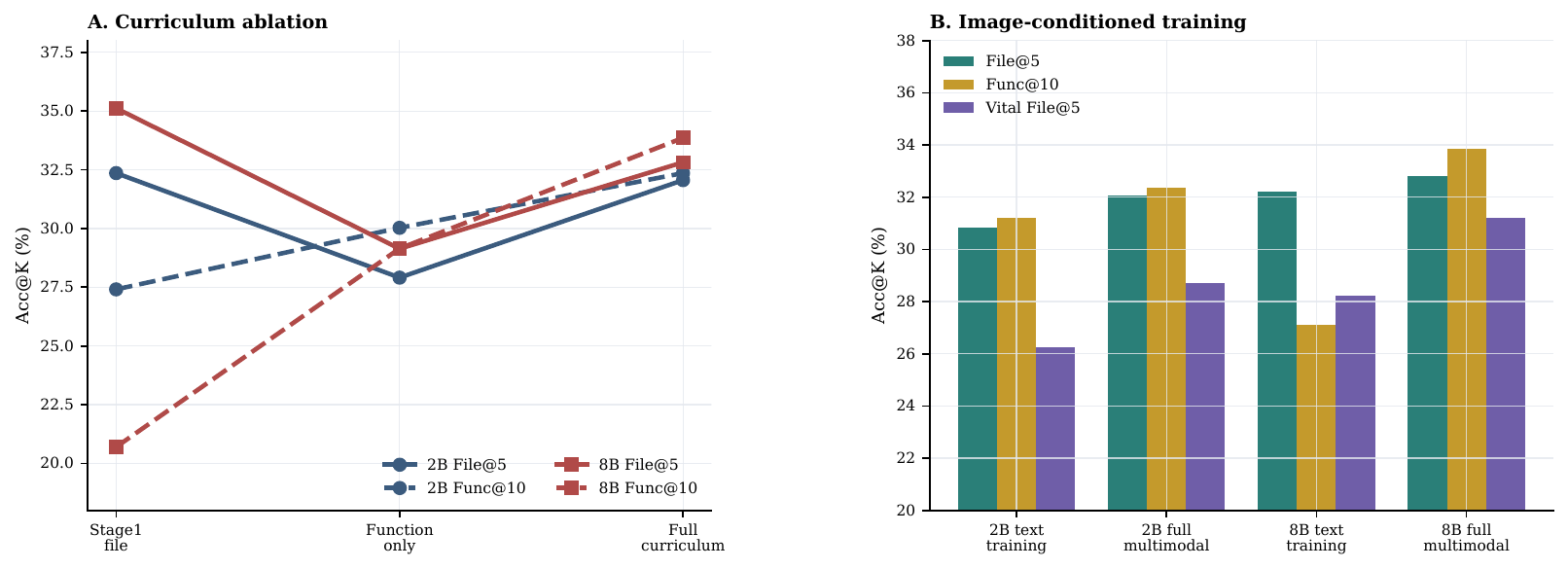}
\caption{Controlled retriever training ablations. The full curriculum improves function ranking over file-only training, and image-conditioned training improves the strongest 8B function and vital-image cells.}
\label{fig:app_training_recipe}
\end{figure}

The main paper uses MM-IssueLoc-VL-Embedding as a controlled probe of image-conditioned retrieval. Figure~\ref{fig:app_training_recipe} and Tables~\ref{tab:app_stage} and~\ref{tab:app_image_training} provide the supporting training ablations. These results are not the central contribution of the benchmark, but they verify that the controlled retriever is sensitive to curriculum and image-conditioned training choices.

\begin{table}[h]
\centering
\small
\setlength{\tabcolsep}{4pt}
\resizebox{\linewidth}{!}{
\begin{tabular}{llrrrr}
\toprule
Backbone & Stage & File@1 & File@3 & File@5 & Func@10 \\
\midrule
2B & stage1-file & 14.42 & 26.53 & 32.36 & 27.41 \\
2B & fn-only & 10.89 & 21.93 & 27.91 & 30.03 \\
2B & fulltrain & 13.19 & 25.61 & 32.06 & 32.36 \\
8B & stage1-file & 15.18 & 29.14 & 35.12 & 20.70 \\
8B & fn-only & 11.96 & 24.08 & 29.14 & 29.15 \\
8B & fulltrain & 13.96 & 27.61 & 32.82 & 33.86 \\
\bottomrule
\end{tabular}}
\caption{Retriever curriculum-stage ablation.}
\label{tab:app_stage}
\end{table}

\begin{table}[h]
\centering
\small
\resizebox{\linewidth}{!}{
\begin{tabular}{lrrrr}
\toprule
Model & File@5 & Func@10 & Vital File@5 & Helpful File@5 \\
\midrule
2B text-only training & 30.83 & 31.20 & 26.24 & 32.28 \\
2B full multimodal & 32.06 & 32.36 & 28.71 & 33.86 \\
8B text-only training & 32.21 & 27.11 & 28.22 & 33.07 \\
8B full multimodal & 32.82 & 33.86 & 31.19 & 33.33 \\
\bottomrule
\end{tabular}}
\caption{Training-data image ablation for MM-IssueLoc-VL-Embedding.}
\label{tab:app_image_training}
\end{table}

\subsection{External Image Ablation on SWE-bench-MM}
\label{sec:app_swebenchmm_image_ablation}

We additionally compare with-image and no-image localization on SWE-bench-MM.
This experiment serves as an external sanity check for RQ2, rather than primary
evidence for visual contribution, because SWE-bench-MM does not provide
controlled image relevance, category, or harmful-image annotations.

Table~\ref{tab:app_swebenchmm_image_ablation} shows that image effects on
SWE-bench-MM are also mixed. Mini-SWE-Agent with Claude-Sonnet-4.6 benefits
substantially from images, while OpenHands changes only slightly. Some methods
even perform better without images, suggesting that current systems do not
consistently exploit the visual channel. This trend is consistent with
MM-IssueLoc, but MM-IssueLoc remains the primary benchmark for RQ2 because it
explicitly controls visual evidence.

\begin{table*}[t]
\centering
\small
\setlength{\tabcolsep}{4pt}
\renewcommand{\arraystretch}{1.08}
\resizebox{0.92\linewidth}{!}{
\begin{tabular}{llrrrr}
\toprule
\textbf{Method} & \textbf{Model} &
\textbf{No-image File@5} &
\textbf{With-image File@5} &
\textbf{$\Delta$ File@5} &
\textbf{With-image Func@10} \\
\midrule
AgentLess & Claude-Sonnet-4.6
& 30.39 & 35.29 & -4.90 & 15.38 \\
OpenHands & Claude-Sonnet-4.6
& 45.10 & 43.14 & +1.96 & 28.85 \\
OpenHands & GPT-5.2
& 41.18 & 42.16 & -0.98 & 32.69 \\
Mini-SWE-Agent & Claude-Sonnet-4.6
& 13.73 & 22.55 & -8.82 & 17.31 \\
MM-IssueLoc-VL-Embedding & 8B
& 29.41 & 28.43 & +0.98 & 13.46 \\
\bottomrule
\end{tabular}}
\caption{External image ablation on SWE-bench-MM. $\Delta$ is no-image minus
with-image File@5, so negative values indicate that removing images hurts.
Values are Acc@K percentages.}
\label{tab:app_swebenchmm_image_ablation}
\end{table*}

\subsection{Cross-benchmark Localization Results}
\label{sec:app_cross_bench}

We further evaluate representative systems on SWE-bench-Lite, SWE-bench-Verified, and SWE-bench-MM. This experiment is used to contextualize MM-IssueLoc rather than as     
primary evidence for RQ2, since only MM-IssueLoc provides controlled image relevance and category annotations.
Since none of these three benchmarks ship with localization gold labels, we follow prior work~\citep{chen2025locagent} and derive file- and function-level localization     
ground truth from the repair patches released with each benchmark. Specifically, we take the files and functions modified by the provided fixing patch as the gold set, 
which allows us to score the localization outputs of each system under the same strict Acc@K protocol used on MM-IssueLoc. 

Table~\ref{tab:app_cross_bench} shows a clear gap
between text-dominant and multimodal localization. On SWE-bench-Lite and SWE-bench-Verified, current agent systems already achieve strong localization performance. For    
example, OpenHands reaches 94.53 File@5 on SWE-bench-Lite and 90.20 File@5 on SWE-bench-Verified. However, the same family of systems drops substantially on SWE-bench-MM,
where the best File@5 is 43.14. This contrast suggests that current systems are relatively mature for text-based SWE localization, but still struggle when localization     
requires interpreting multimodal issue evidence. Thus, SWE-bench-MM provides additional evidence that multimodal repository-level localization remains underdeveloped, while
MM-IssueLoc offers the controlled annotations needed to diagnose this gap.

\begin{table*}[t]
\centering
\small
\setlength{\tabcolsep}{3.5pt}
\renewcommand{\arraystretch}{1.08}
\resizebox{0.8\linewidth}{!}{
\begin{tabular}{llrrrrr}
\toprule
\textbf{Benchmark} & \textbf{Method / Model} &
\textbf{File@1} & \textbf{File@3} & \textbf{File@5} &
\textbf{Func@5} & \textbf{Func@10} \\
\midrule
\multicolumn{7}{l}{\textbf{SWE-bench-Lite}} \\
\cmidrule(lr){1-7}
 & AgentLess / Claude-Sonnet-4.6
& 78.47 & 89.42 & 91.24 & 70.44 & 70.80 \\
 & OpenHands / Claude-Sonnet-4.6
& \textbf{87.59} & \textbf{94.16} & 94.16 & \textbf{79.20} & \textbf{79.20} \\
 & OpenHands / GPT-5.2
& 83.21 & 93.43 & \textbf{94.53} & 74.09 & 74.45 \\
 & Mini-SWE-Agent / Claude-Sonnet-4.6
& 72.99 & 73.72 & 73.72 & 63.87 & 63.87 \\
\midrule
\multicolumn{7}{l}{\textbf{SWE-bench-Verified}} \\
\cmidrule(lr){1-7}
 & AgentLess / Claude-Sonnet-4.6
& 69.60 & 84.40 & 86.80 & 51.45 & 52.34 \\
 & OpenHands / Claude-Sonnet-4.6
& \textbf{77.80} & 87.40 & 88.20 & \textbf{59.47} & \textbf{59.69} \\
 & OpenHands / GPT-5.2
& 74.40 & \textbf{88.40} & \textbf{90.20} & 57.02 & 57.68 \\
 & Mini-SWE-Agent / Claude-Sonnet-4.6
& 65.60 & 68.80 & 68.80 & 49.44 & 49.44 \\
\midrule
\multicolumn{7}{l}{\textbf{SWE-bench-MM with images}} \\
\cmidrule(lr){1-7}
 & AgentLess / GPT-5.2
& 14.71 & 28.43 & 33.33 & 15.38 & 17.31 \\
 & OpenHands / Claude-Sonnet-4.6
& \textbf{29.41} & \textbf{41.18} & \textbf{43.14} & 28.85 & 28.85 \\
 & OpenHands / GPT-5.2
& 24.51 & 37.25 & 42.16 & \textbf{32.69} & \textbf{32.69} \\
 & Mini-SWE-Agent / Claude-Sonnet-4.6
& 18.63 & 22.55 & 22.55 & 17.31 & 17.31 \\
 & MM-IssueLoc-VL-Embedding / 8B
& 16.67 & 23.53 & 28.43 & 11.54 & 13.46 \\
\bottomrule
\end{tabular}}
\caption{Cross-benchmark localization results. SWE-bench-MM is evaluated with
images, while SWE-bench-Lite and SWE-bench-Verified are text-only benchmarks.
Values are Acc@K percentages.}
\label{tab:app_cross_bench}
\end{table*}

\section{Benchmark Details}
\label{app:benchmark-details}

\subsection{Released Views and Statistics}
\label{app:benchmark_statistics}

MM-IssueLoc contains 652 canonical instances. The file-level view contains all 652 rows, and the function-level view contains 343 rows with non-empty extracted edit functions. The annotation split is 450 human and 202 AI-assisted rows. The difficulty split is 214 easy, 263 medium, and 176 hard instances. The largest language groups are TypeScript (151), Python (126), JavaScript (120), C++ (45), Java (44), Go (33), C\# (33), Rust (27), C (21), and PHP (20), with 23 languages in total.

\begin{table}[h]
\centering
\small
\resizebox{1.0\linewidth}{!}{
\begin{tabular}{lrrrrrrr}
\toprule
Category & UI & Behavior & Error & Rendering & Code & Log & Data-viz \\
\midrule
Instances & 177 & 99 & 92 & 85 & 84 & 66 & 50 \\
\bottomrule
\end{tabular}}
\caption{Image category distribution. Categories are mutually exclusive at the instance level.}
\label{tab:app_category}
\end{table}

\begin{table}[h]
\centering
\small
\resizebox{1.0\linewidth}{!}{
\begin{tabular}{lrrrr}
\toprule
Relevance & Vital (+2) & Helpful (+1) & Neutral (0) & Harmful (-1) \\
\midrule
Instances & 202 & 381 & 14 & 55 \\
\bottomrule
\end{tabular}}
\caption{Per-image relevance distribution used for stratified evaluation.}
\label{tab:app_relevance}
\end{table}

\subsection{Benchmark Examples}
\label{app:benchmark-examples}

\begin{figure*}[t]
  \centering
  \includegraphics[width=0.9\linewidth]{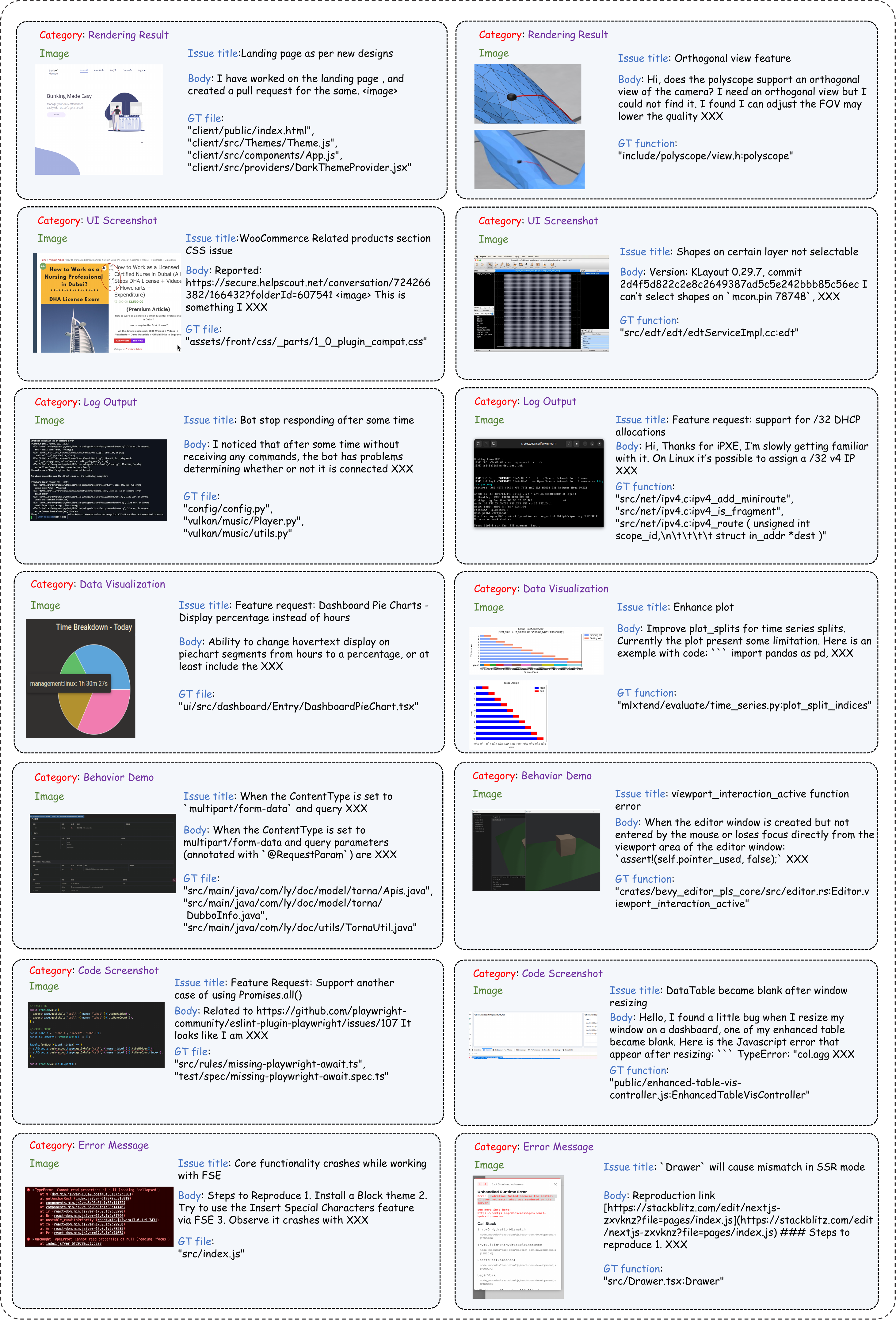}
  \caption{Representative MM-IssueLoc examples grouped by image category.}
  \label{app:benchmark_examples_with_category}
\end{figure*}

\Cref{app:benchmark_examples_with_category} presents representative
examples from MM-IssueLoc, organized by the image categories defined in~\Cref{tab:app_category}. Each example illustrates the issue context,
the associated visual evidence, and the corresponding gold annotations at both
file and function levels.

\subsection{AI Annotation Gate}
\label{app:ai_annotation}
The AI-assisted extension uses four stages. Stage A estimates relevance and admits only +1 or +2 rows for the main augmented subset. Stage B assigns one of seven image categories with confidence at least 0.6. Stage C judges whether the proposed patch plausibly addresses the issue. Stage D repeats relevance and category with a second VLM and accepts only if relevance differs by at most one level and category matches. The accepted rows are marked with \texttt{annotation\_by=ai}.

\subsection{Harmful-Image Synthesis}
\label{app:harmful-synthesis}

The harmful subset uses same-repository image-text mismatches, TF-IDF-similar issues with disjoint edit files, and comment-thread image promotion. Each candidate receives a seven-axis quality assessment: text sufficiency, surface relevance, image direction, direction deviation, misleading score, developer impact, and suggested label. Human review filters implausible or insufficiently misleading candidates.

\section{Visual Content Evidence Extraction}
\label{app:vce}

\subsection{VCE Mechanism}
\label{app:vce-mechanism}

Visual Content Evidence (VCE) converts issue images into structured textual
evidence before localization. Instead of directly using the raw image as model
input, VCE extracts task-relevant visual content, such as error messages, UI
states, code snippets, logs, rendering artifacts, and chart patterns, into a
compact textual representation. This design provides a diagnostic side channel
for separating the contribution of visual content from the effect of direct
image conditioning. In our evaluation, VCE is used to test whether localization
improvements come from information contained in the image or from the model's
ability to process the image modality itself.

\begin{figure*}[t]
  \centering
  \includegraphics[width=0.95\linewidth]{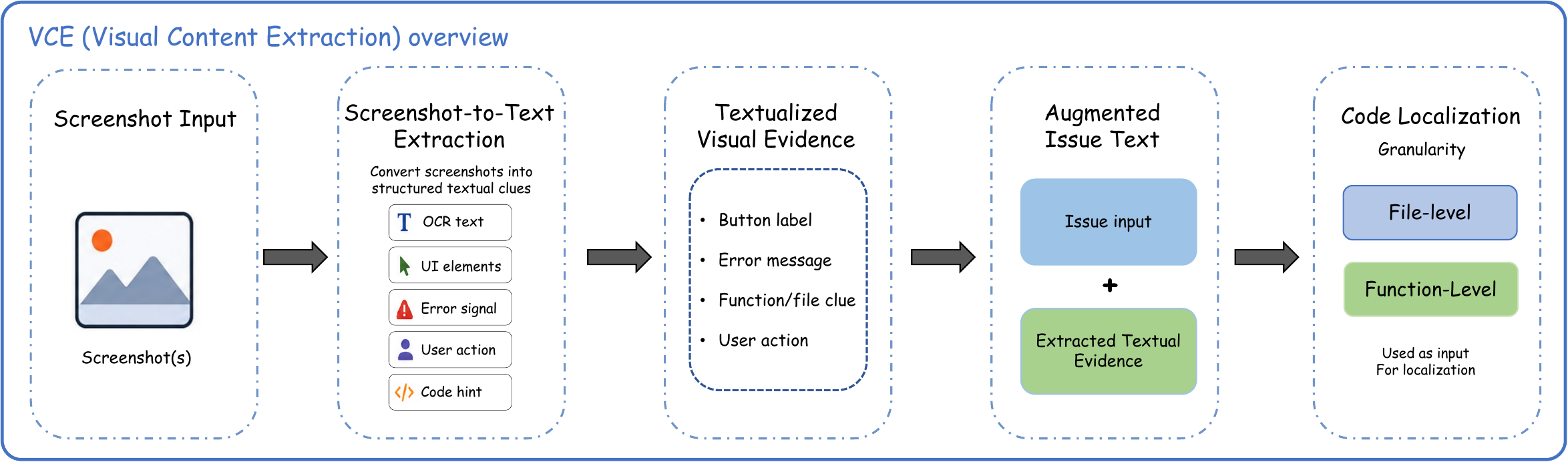}
  \caption{Overview of the VCE extraction mechanism. Issue images are converted
  into structured textual evidence and then used as an auxiliary diagnostic
  input for issue localization.}
  \label{app:vce_method_overview}
\end{figure*}

\subsection{VCE Extraction Examples}
\label{app:vce-examples}

\begin{figure*}[ht]
  \centering
  \includegraphics[width=0.96\linewidth]{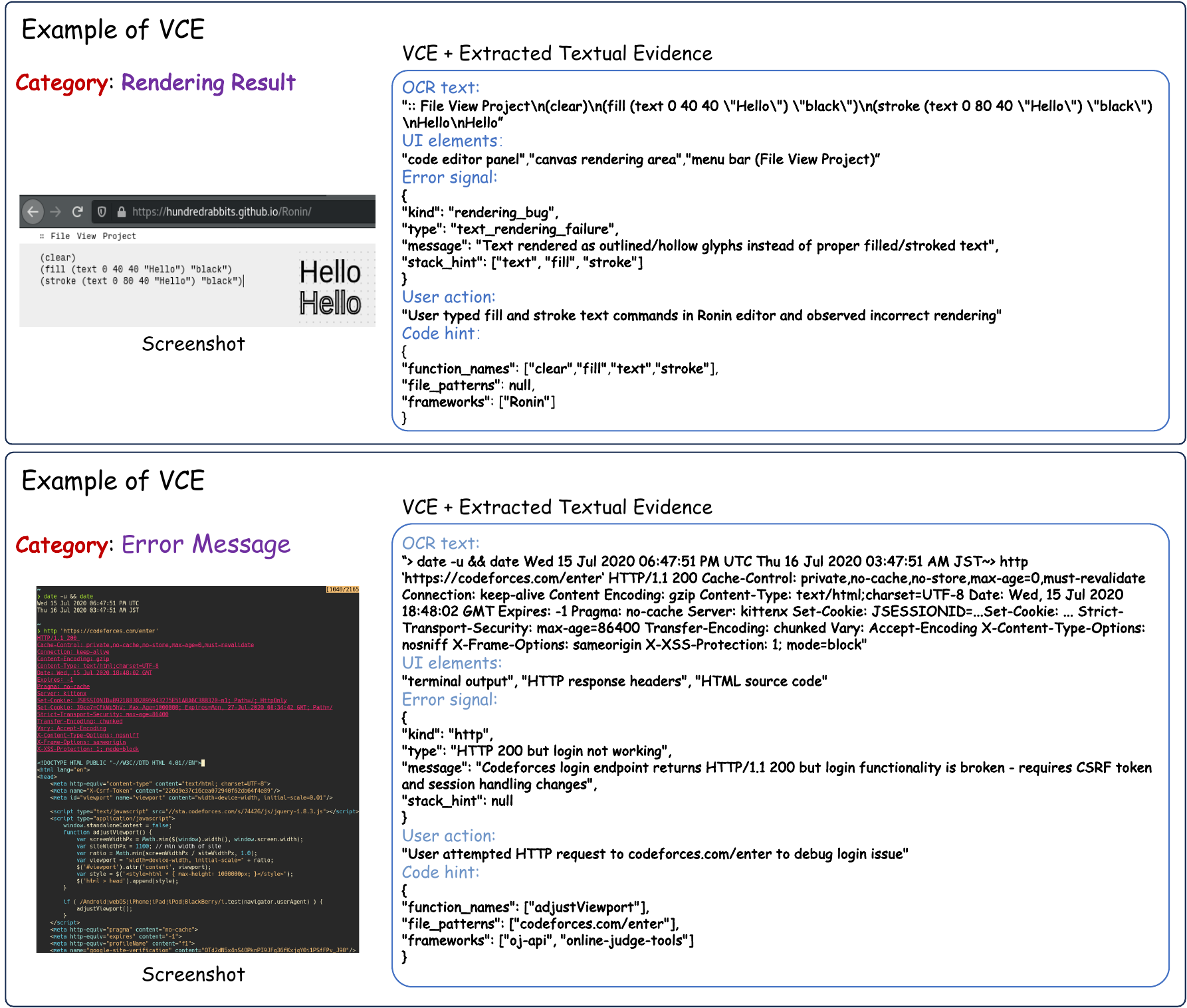}
  \caption{Representative VCE extraction examples across image categories,
  including error messages, UI screenshots, behavior demos, rendering results,
  code screenshots, data visualizations, and log outputs.}
  \label{app:vce_example}
\end{figure*}

Figure~\ref{app:vce_example} shows representative image-to-VCE extraction
examples across different image categories. Each example contains the original
visual evidence, the extracted VCE text, and its corresponding interpretation
for file-level and function-level localization.

\subsection{VCE Prompt}
\label{app:vce-prompt}
\begin{figure*}[p]
\centering
\includegraphics[
  width=1.0\linewidth,
  height=0.95\textheight,
  keepaspectratio
]{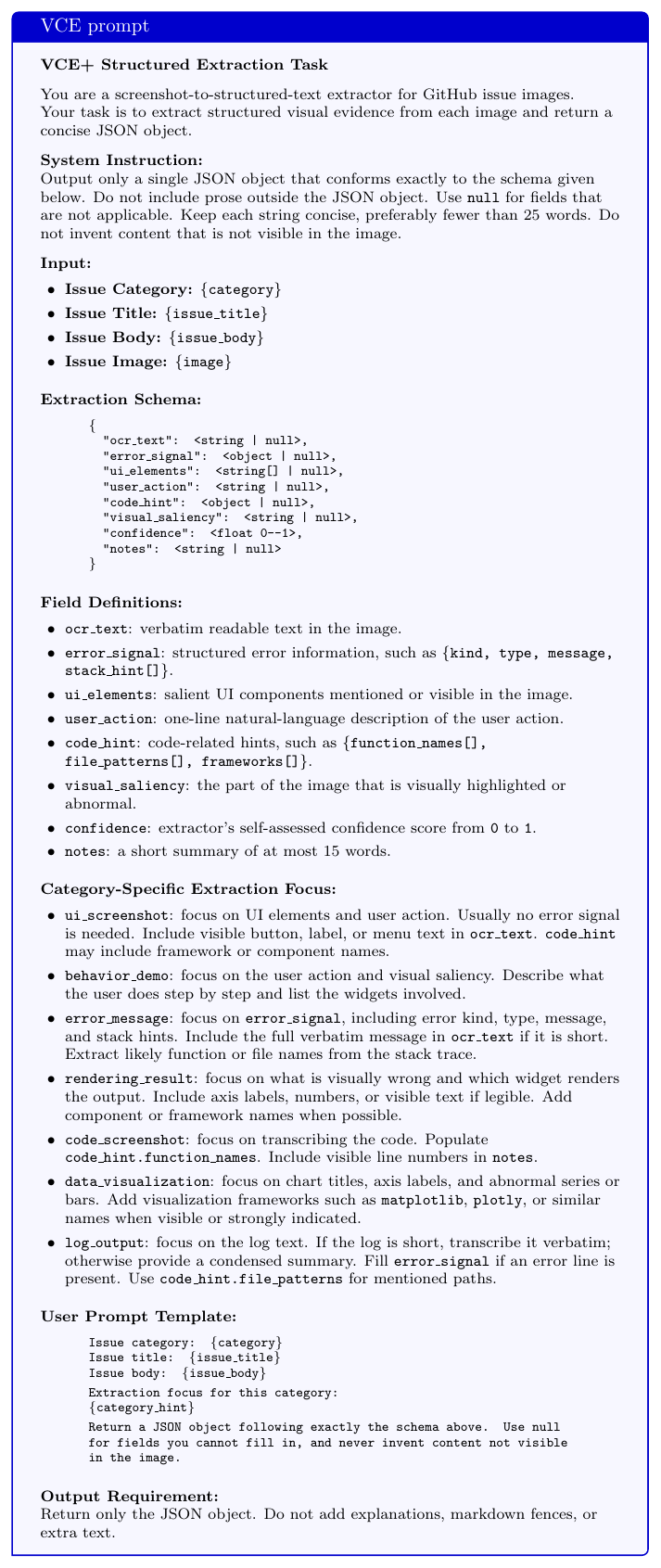}
\caption{VCE prompt}
\label{app:vce_prompt}
\end{figure*}

The full prompt template is provided in Figure~\ref{app:vce_prompt}.

\section{Prompt Templates}
\label{app:prompts}

\subsection{AI-Driven Harmful Impact Prompt}
\label{app:prompt-harmful-impact}
\begin{figure*}[p]
\centering
\includegraphics[
  width=1.0\linewidth,
  height=0.95\textheight,
  keepaspectratio
]{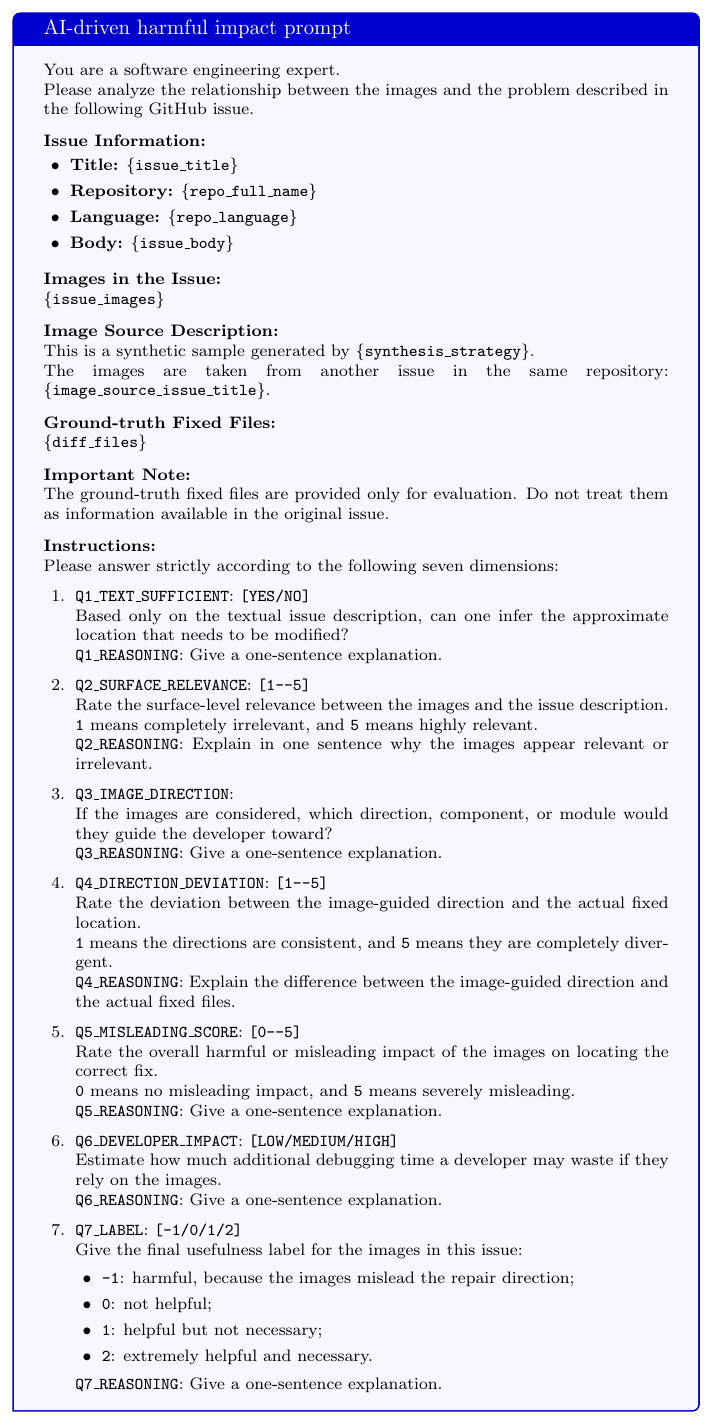}
\caption{AI-Driven Harmful Impact Prompt}
\label{app:AI-driven harmful impact prompt}
\end{figure*}

The full prompt template is provided in Figure~\ref{app:AI-driven harmful impact prompt}.

\subsection{AgentLess Prompt}
\label{app:prompt-AgentLess}
\begin{figure*}[p]
\centering
\includegraphics[
  width=1.0\linewidth,
  height=0.95\textheight,
  keepaspectratio
]{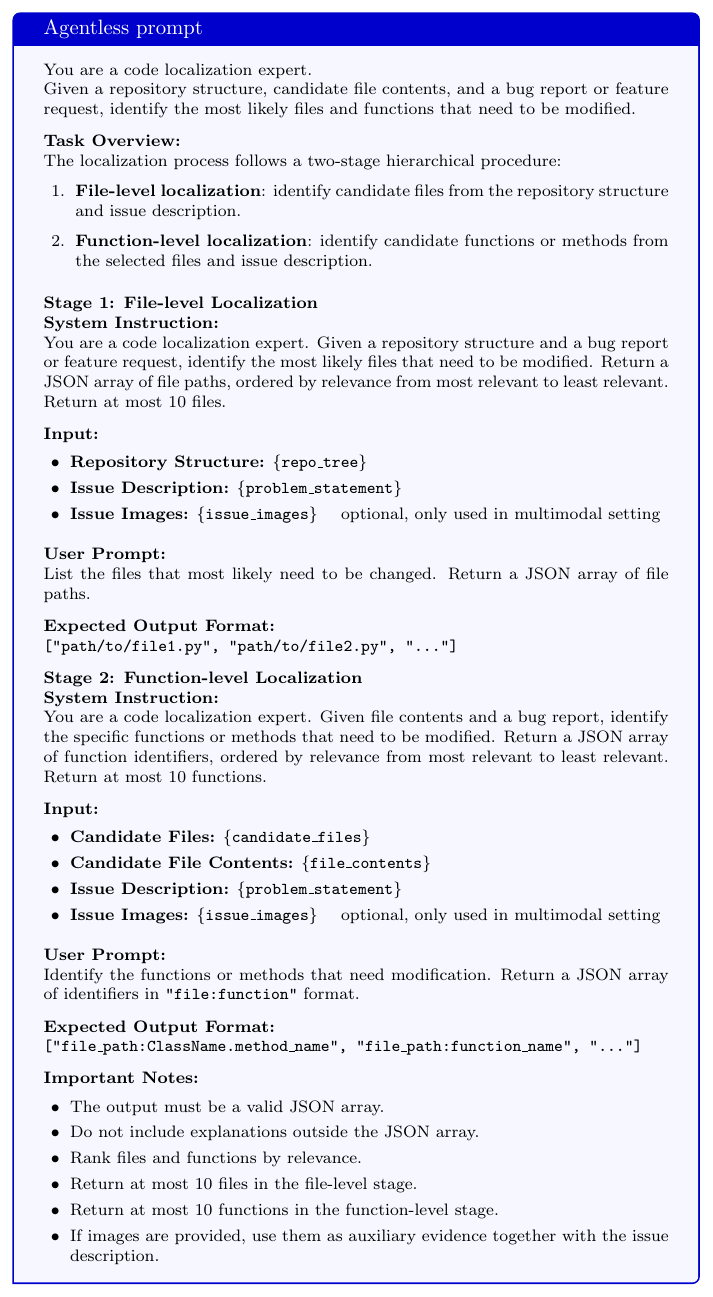}
\caption{AgentLess Prompt}
\label{app:AgentLess Prompt}
\end{figure*}

The full prompt template is provided in Figure~\ref{app:AgentLess Prompt}.

\subsection{LocAgent Prompt}
\label{app:prompt-locagent}

\begin{figure*}[p]
\centering
\includegraphics[
  width=1.0\linewidth,
  height=0.95\textheight,
  keepaspectratio
]{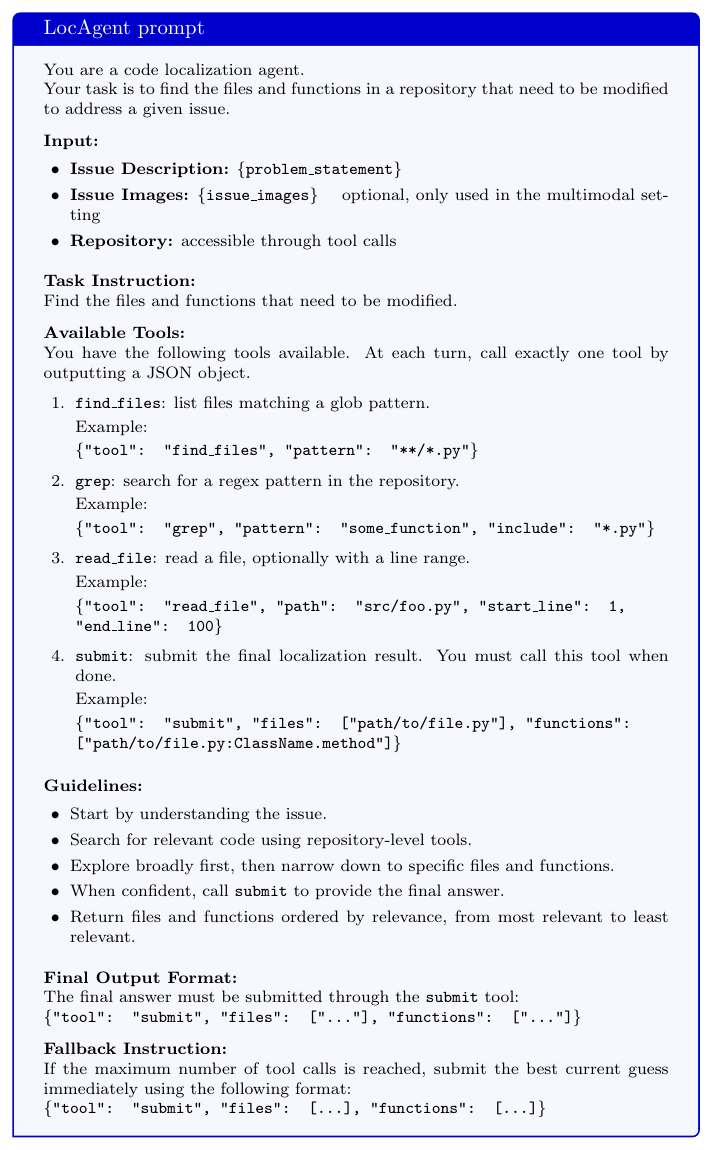}
\caption{LocAgent prompt}
\label{app:LocAgent prompt}
\end{figure*}

The full prompt template is provided in Figure~\ref{app:LocAgent prompt}.

\subsection{OpenHands Prompt}
\label{app:prompt-openhands}
\begin{figure*}[p]
\centering
\includegraphics[
  width=1.0\linewidth,
  height=0.95\textheight,
  keepaspectratio
]{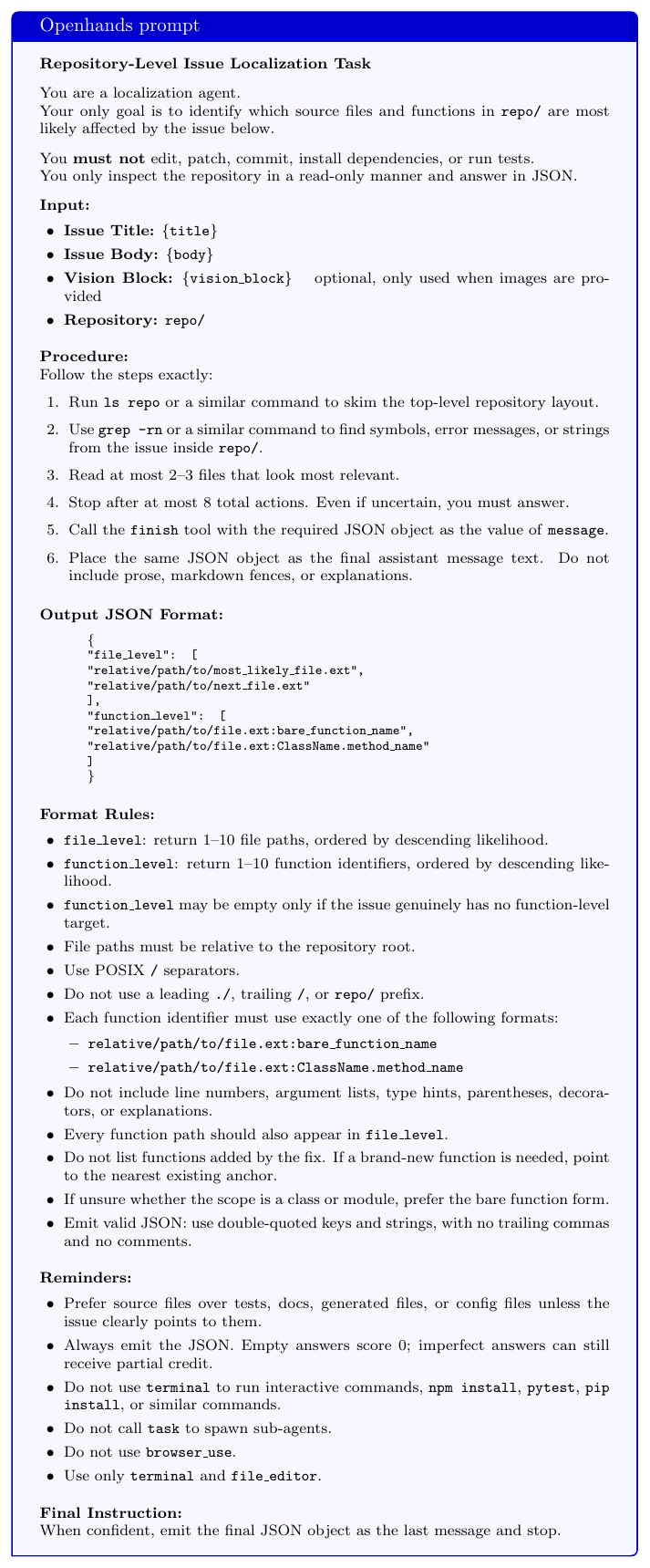}
\caption{Openhands prompt}
\label{app:Openhands prompt}
\end{figure*}

The full prompt template is provided in Figure~\ref{app:Openhands prompt}.

\subsection{Mini-SWE-Agent Prompt}
\label{app:prompt-Mini-SWE-Agent}
\begin{figure*}[p]
\centering
\includegraphics[
  width=1.0\linewidth,
  height=0.95\textheight,
  keepaspectratio
]{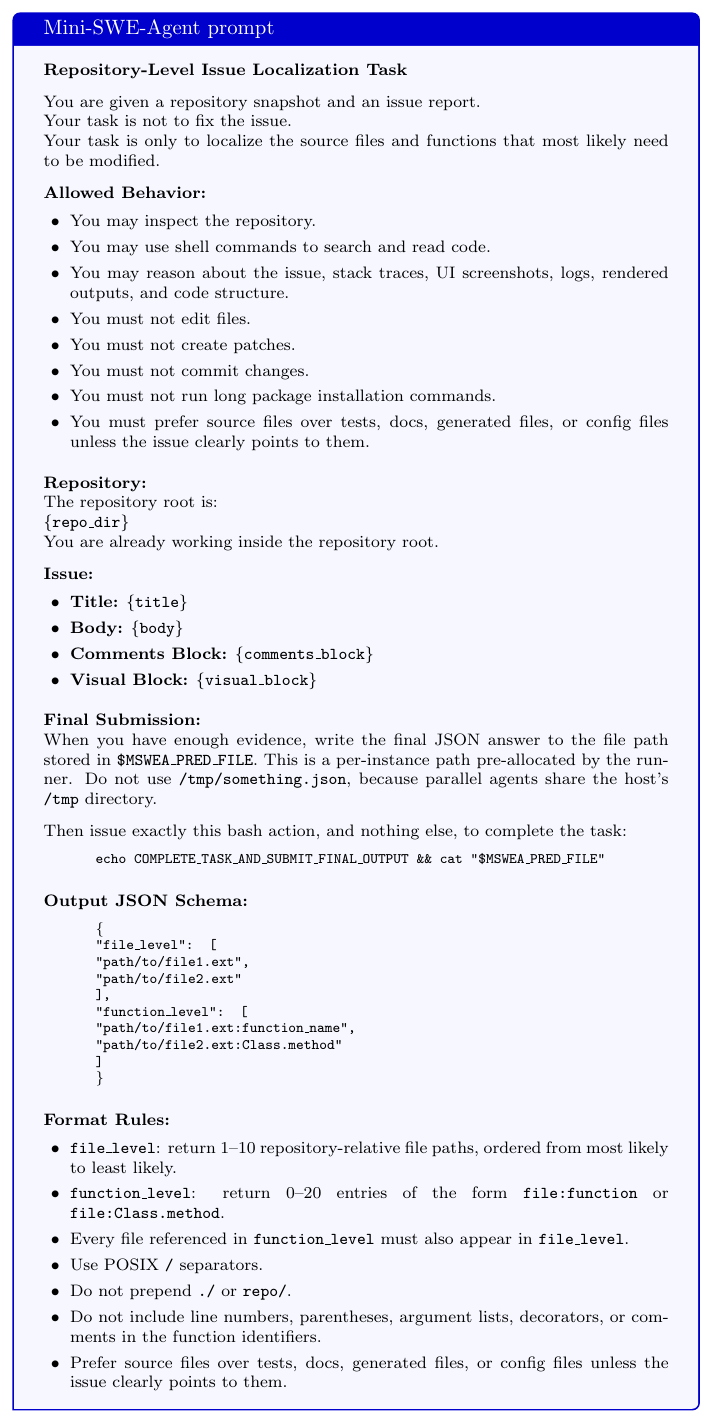}
\caption{Mini-SWE-Agent Prompt}
\label{app:Mini-SWE-Agent Prompt}
\end{figure*}

The full prompt template is provided in Figure~\ref{app:Mini-SWE-Agent Prompt}.

\end{document}